# From Loss Diagnosis to Rational Design: A Unified Analytical Model for Photoelectrochemical Cells

Ziyan Pan, Giulia Tagliabue*
Laboratory of Nanoscience for Energy Technologies (LNET), Institute of Mechanical Engineering, School of Engineering, École Polytechnique Fédérale de Lausanne (EPFL), CH-1015 Lausanne, Switzerland
*Corresponding author: giulia.tagliabue@epfl.ch

## Abstract

Photoelectrochemical (PEC) cells are a compelling route to solar-driven chemical energy storage and feedstock synthesis, yet their deployment is hindered by coupled losses spanning light absorption, carrier transport, interfacial charge transfer, and semiconductor–electrolyte matching. Existing models address these losses in an architecture-specific manner and fall short of quantitative experimental diagnosis or actionable design guidance. Here, we introduce a unified loss-analysis framework applicable to both built-in junction (BIJ) and semiconductor–electrolyte junction (SEJ) photoelectrodes within a consistent set of physically meaningful parameters. The framework delivers current–voltage curves and efficiency metrics under ideal and real conditions, constructing efficiency maps to delineate theoretical limits and material-selection windows. Critically, by fitting experimental current–voltage data, it enables quantitative energy-loss decomposition into thermodynamic, optical, recombination, and interfacial contributions, pinpointing performance bottlenecks in real devices and mapping them directly onto optimization strategies such as co-catalyst integration or nanostructuring. Energy flows are visualized through Sankey diagrams, providing an intuitive picture of how incident solar energy is absorbed, dissipated, or converted into chemical output. Validated against state-of-the-art literature results spanning solar water splitting, $CO_2$ reduction, $NH_3$ synthesis, and solar redox flow batteries, the framework further enables systematic comparison of photovoltaic-grade absorbers (e.g., Si, perovskites) with intermediate-bandgap semiconductors (e.g., $\alpha$-$Fe_2O_3$, $BiVO_4$), identifying the key factors limiting each material class. Together, these capabilities support a paradigm shift from empirical optimization to mechanism-informed rational design of high-efficiency PEC energy-conversion systems.

## 1. Introduction

With the rapid growth in global energy demand and escalating climate challenges[1,2], developing a low-carbon energy system remains a central objective of the global energy transition. Solar energy, characterized by its abundance, wide distribution, and near-inexhaustibility, is widely regarded as one of the most promising energy sources[3]. However, its intermittency limits reliable energy supply, necessitating efficient strategies for solar energy storage. Converting solar energy into stable chemical bonds offers a viable pathway toward dispatchable utilization[4].

Currently, solar-to-chemical conversion is primarily realized through photovoltaic–electrolysis, photoelectrochemical (PEC), and photocatalytic systems[5]. Among these, PEC systems integrate light absorption and electrochemical reactions within a single device, enabling direct solar-to-chemical energy conversion. They further offer tunable interface reactions and facile product separation, making them particularly attractive for solar energy storage[6]. In recent years, PEC systems have been extensively explored for water splitting[7], $CO_2$ reduction[4], solar redox flow batteries[8], and value-added chemical synthesis[9].

Despite their promise, PEC systems remain limited by low solar-to-chemical conversion efficiency, which is a key barrier to large-scale deployment[10]. Efficiency losses originate from multiple physical and chemical processes, including insufficient light absorption and charge carrier recombination in semiconductors, sluggish interface charge transfer kinetics and high overpotentials, as well as band structure alignment with redox potentials and cell architecture design[11–13]. These processes are intrinsically coupled, making it challenging to isolate and quantify individual loss mechanisms experimentally. Therefore, a quantitative framework that integrates carrier dynamics and interface charge transfer with current-voltage analysis is essential for disentangling loss pathways, elucidating PEC operation, and enabling targeted system optimization.

Since the 1950s, theoretical modeling of PEC systems has continuously evolved, leading to a range of frameworks for describing the current-voltage (J-V) characteristics of photoelectrodes. Based on the mechanisms of photogenerated carrier separation, photoelectrodes can be broadly classified into two categories. The first relies on solid-state junctions within the semiconductor to drive carrier separation, referred to as built-in junction (BIJ) photoelectrodes (Fig. 1(a)). The second relies on Fermi level equilibration at the semiconductor-electrolyte interface, which induces band bending and the associated space-charge electric field, and is referred to as semiconductor-electrolyte junction (SEJ) photoelectrodes (Fig. 1(b)).

For BIJ photoelectrodes, as charge carriers are generated and separated within the semiconductor, bulk semiconductor processes can be decoupled from interface reaction kinetics. Carrier dynamics in the semiconductor are typically described using classical p-n junction theory, while interface charge transfer is modeled using the Butler-Volmer equation. On this basis, Shaner et al. developed a corresponding J-V model[14], which was further applied by Fountaine et al. to PEC water-splitting systems to evaluate theoretical efficiency limits[15]. Related thermodynamic and multiphysics simulations have also been reported by Gokhale et al., Qureshy et al., and Jin et al.[16–18]. We note that such models are often directly applied to SEJ photoelectrodes. However, in SEJ systems, the semiconductor is in direct contact with the electrolyte and carrier separation occurs at the interface. Therefore, the decoupling approximation used in BIJ fails to capture the underlying physics of SEJ and may introduce systematic deviations.

For SEJ photoelectrodes, early studies established a systematic theoretical foundation. Gärtner, Butler, and Gerischer proposed frameworks describing carrier generation, recombination, and transport in semiconductors, leading to the development of the Gärtner-Butler model[19–21]. Subsequently, Lindquist et al. and Sodergren et al. applied these models to analyze quantum efficiency of photoelectrodes[22,23]. In the past decade, Bedoya-Lora et al. and Falciani et al. have introduced bulk and surface loss parameters to refine these models[24,25]. However, these approaches often overlook the fundamental differences in interface charge transfer mechanisms between semiconductor and metal electrodes, adopting Butler-Volmer kinetics in a simplified manner. To address this limitation, Reiss, Chazalviel and Vanmaekelbergh developed interface charge transfer theories specifically for semiconductor electrodes and derived J-V characteristics under dark conditions[26–28]. Building on this, Reichman, as well as Nozik and Memming coupled semiconductor carrier transport theory with Marcus-Gerischer interface charge transfer theory, establishing SEJ models that more faithfully represent the underlying physical and chemical processes[29,30]. In parallel, Chen et al., Gaudy et al. and Mills et al. adopted modified Schottky junction approximations to simplify semiconductor-electrolyte interfaces, enabling tractable mathematical modeling and multiphysics simulations[31–33].

Overall, despite substantial progress in PEC modeling, existing theoretical frameworks remain fragmented: multiple approaches coexist, yet a unified description of the dominant loss mechanisms across both BIJ and SEJ architectures within a consistent set of physical parameters is still lacking. Furthermore, current analytical models predominantly target theoretical efficiency-limit calculations, and a systematic framework for quantifying the contribution of each individual loss mechanism in experimentally realized devices is still underdeveloped. Such a framework is an essential step toward guiding material selection and system design for targeted improvements.

To address these limitations, this work introduces a unified loss-analysis framework for the quantitative modeling and diagnosis of PEC systems. The model defines a consistent set of physically meaningful parameters capturing non-ideal light absorption, carrier transport and recombination, interfacial charge transfer, and junction energetics. Embedded into analytical J–V expressions, these parameters allow SEJ and BIJ photoelectrodes to be described within a common loss formalism: the SEJ model extends Reichman's treatment of carrier generation, transport, and interfacial charge transfer, whereas the BIJ model refines the formulation proposed by Shaner et al. This approach enables consistent calculation of theoretical and practical efficiency limits across junction architectures.

Beyond efficiency limits, the framework can fit experimental J–V data to decompose efficiency losses into thermodynamic, optical absorption, recombination, and interfacial contributions, thereby identifying the dominant bottlenecks in real devices. The resulting energy flows are visualized using Sankey diagrams, providing an intuitive picture of how incident solar energy is absorbed, dissipated, or converted into chemical output, guiding towards targeted improvement strategies, such as co-catalyst integration or nanostructuring. The framework is validated through extensive comparison with state-of-the-art literature results spanning solar water splitting, $CO_2$ reduction, $NH_3$ synthesis, and solar redox flow batteries, demonstrating its applicability across a broad range of energy-relevant chemistries. Furthermore, the framework is used to compare devices based on photovoltaic-grade absorbers (e.g., Si and perovskites) with those based on commonly used, non-photovoltaic-grade intermediate-bandgap semiconductors (e.g., $\alpha$-$Fe_2O_3$ and $BiVO_4$), enabling the identification of key factors limiting performance in each material class and providing guidance for further optimization. Overall, the framework provides a unified tool to diagnose loss mechanisms, identify performance bottlenecks, determine efficiency limits, guide material selection and interface matching, and evaluate PEC materials across multiple dimensions. Together, these capabilities support a paradigm shift from empirical optimization to mechanism-informed rational design of high-efficiency PEC energy-conversion systems.

# 2. A Unified Framework for Performance Analysis and Loss Decomposition in PEC Systems

An instructive PEC model should satisfy the following key requirements: (i) accurately capture the essential physical and chemical processes in PEC systems, including charge generation, recombination, and transport in semiconductors, as well as interface charge transfer at the electrode-electrolyte interface; (ii) reproduce experimental observations, ensuring the reliability and testability of model predictions; (iii) exhibit sufficient generality to be applicable across diverse PEC systems, such as water splitting, $CO_2$ reduction, $NH_3$ synthesis, and solar redox flow batteries; (iv) remain simple and tractable, avoiding excessive parameters and computational complexity; (v) enable the identification and quantitative analysis of performance bottlenecks, particularly through the analysis of energy loss pathways; (vi) provide actionable guidance for device design, including the selection of photoelectrode materials and the matching of electrode-electrolyte systems.

This Section presents the analytical expressions for the J–V curves of BIJ and SEJ photoelectrodes, together with the unified loss-descriptors. On this basis, efficiency maps are constructed to delineate the performance limits of ideal and practical devices. The contribution of each loss pathway is then quantified to pinpoint the dominant bottlenecks of practical devices, and a practical toolkit is provided for their optimization. Section 3 applies this framework to validate the model against experimental data and to analyze representative PEC materials and devices, establishing their theoretical limits and practical performance upper bounds, diagnosing the dominant performance bottlenecks and corresponding improvement directions, and comparing different material systems to inform future device design. Finally, Section 4 discusses the strengths and limitations of the model.

## 2.1. Description of Photoelectrode J–V Characteristics and Unified Model Construction

PEC systems generate electrons and holes in light-absorbing materials and transport the photogenerated carriers to the photoelectrode–electrolyte interface to drive target redox reactions, as illustrated in Fig. 1(c). The overall performance of PEC devices is ultimately reflected in the relationship between photocurrent density and the applied voltage, namely the J–V characteristics. Therefore, the J–V relationship is not only the fundamental representation of device output performance, but also the central framework for quantitatively linking physical loss mechanisms to performance metrics.

At present, PEC systems can be broadly categorized into two representative architectures (Fig. 1(a,b)): BIJ PEC devices, which are typically based on high-quality photovoltaic-grade materials, and SEJ PEC devices, which generally employ low-cost, earth-abundant, and medium bandgap semiconductor materials. Although these two types of devices differ significantly in their junction structures and operating mechanisms, their performance losses can be attributed to a common set of underlying physical origins.

As shown in Fig. 1(c), these losses include thermodynamic losses as described by the Shockley–Queisser limit[34], incomplete light absorption and carrier recombination losses within the semiconductor, as well as parasitic circuit losses and interface charge-transfer kinetic losses associated with the interface.

Under ideal conditions, the semiconductor reverse saturation current, typically labelled as $J_{sc}^{0,ideal}$ in BIJ devices and $J_{h}^{0,ideal}$ in SEJ devices, is determined by the Shockley–Queisser detailed balance principle[34]:

$$J_{sc}^{0,ideal}(\text{or } J_{h}^{0,ideal})=\frac{q}{4\pi^2\hbar^3c^2}\cdot\int_{E_g}^{\infty}\frac{E^2}{e^{\frac{E}{kT}}-1}\cdot dE \tag{1}$$

where $\hbar$ is the reduced Planck constant, c is the speed of light, $E_g$ is the semiconductor bandgap, k is the Boltzmann constant and T is the temperature. In practical devices, non-radiative recombination of photogenerated charges leads to a reverse saturation current higher than the ideal value. A recombination loss factor $\beta\in[0,1]$ can thus be introduced to describe the real reverse saturation current as a function of the ideal one:

$$J_{sc}^{0}=\frac{J_{sc}^{0,ideal}}{\beta}\ (\text{or } J_{h}^{0}=\frac{J_{h}^{0,ideal}}{\beta}) \tag{2}$$

Concurrently, the ideal photocurrent density $J_{L,ideal}$ can be obtained from the AM1.5G solar spectrum as:

$$J_{L,ideal}=q\cdot\int_{0}^{\lambda_g}\frac{E_{AM1.5G}(\lambda)\cdot\lambda}{hc}\cdot d\lambda \tag{3}$$

where q is the elementary charge, $\lambda_g$ is the cutoff wavelength corresponding to the bandgap, and $E_{AM1.5G}(\lambda)$ is the incident solar spectral irradiance. In real devices, deviations between actual optical absorption and carrier conversion efficiency from the ideal case must be considered. A current loss factor $\theta\in[0,1]$ is thus introduced to quantify the real-device photocurrent as:

$$J_L=\theta J_{L,ideal} \tag{4}$$

Based on these shared loss mechanisms, the J–V relationships for BIJ and SEJ photoanodes can be formulated as follows (see Supplementary Information Section 1 for detailed derivation and photocathode expressions):

$$f_{BIJ}(J,V)=V+JR_s-\frac{kT}{q}\ln\left(\frac{J_L-(J+\frac{V+JR_s}{R_{sh}})}{J_{sc}^{0}}+1\right)+\frac{2kT}{q}\text{arcsinh}\left(\frac{J+\frac{V+JR_s}{R_{sh}}}{2J_{ex}}\right)=0 \tag{5}$$

$$f_{SEJ}(J,V)=J+\frac{V+JR_s}{R_{sh}}-\frac{J_h^0(1-e^{\frac{q(V+JR_s)}{kT}})+J_L}{1+\frac{J_h^0}{J_{ex,h}}e^{\frac{q(V+JR_s)}{kT}}}+J_{ex,e}(e^{\frac{q(V+JR_s)}{kT}}-1)=0 \tag{6}$$

where $R_s$ is the series resistance, $R_{sh}$ is the shunt resistance and $J_{ex}$ is the exchange current density at the interface (BIJ) while $J_{ex,h}$ and $J_{ex,e}$ represent the interface exchange current densities for holes and electrons in SEJ, respectively. If the details of the chemical reaction are fully known, including the reaction rate constant, the density of states of the reduced and oxidized species as well as the reorganizational energy, $J_{ex(e,h)}$ can be calculated theoretically (see Supplementary Information Section 1 and Section 2). However, these quantities are difficult to obtain experimentally. In the following, $J_{ex(e,h)}$ is therefore treated as a parameter that reflects the charge transfer losses at the solid-liquid interface. In the ideal case $J_{ex,h}\rightarrow\infty$, $J_{ex,e}\rightarrow 0$ and all charge carriers reaching the interface participate in a chemical reaction with zero overpotential.

The different J–V expressions obtained for BIJ and SEJ photoelectrodes within the unified loss-parameter framework reflect the distinct ways in which these losses affect device performance. Nonetheless, it is critical to realize that the two formulations reduce to the same ideal diode equation under perfect absorption ($\theta=1$) as well as loss-free charge transport ($\beta=1$, $R_s=0$) and charge transfer ($J_{ex}\rightarrow\infty$, $J_{ex,h}\rightarrow\infty$, $J_{ex,e}\rightarrow 0$, $R_{sh}\rightarrow\infty$) conditions:

$$V=\frac{kT}{q}\ln\left(\frac{J_{L,ideal}-J}{J_{sc}^{0,ideal}(\text{or } J_{h}^{0,ideal})}+1\right) \tag{7}$$

Based on these relationships, performance metrics such as the short-circuit current $J_{SC}$, open-circuit voltage $V_{OC}$, photoelectrode efficiency $\eta_{PE}$, and fill factor FF can be further defined by analogy with the corresponding definitions used for solar cells, enabling quantitative evaluation of photoelectrode performance (see Supplementary Information Section 3 for detailed expressions).

Finally, for a complete PEC system, the actual operating point ($J_{OP}$, $V_{OP}$) is determined by the intersection of the J–V curves of the photoelectrode and the counter electrode. Using the Butler-Volmer equation to describe the J–V relationship of the metal counter electrode and defining $\Delta U$ as the reaction potential difference of the counter electrode relative to the photoelectrode side (see Supplementary Information Section 3), the overall energy conversion efficiency of the PEC cell (PEC efficiency) is defined as:

$$\eta_{PEC}=\frac{J_{OP}\cdot\Delta U}{P_{in}}=\gamma\eta_{PE} \tag{8}$$

where $P_{in}$ is the incident solar power density and $\gamma$ is the utilization factor of the photoelectrode, which quantifies how effectively its intrinsic output capability is utilized under operating conditions.

Overall, using the above expressions, the performance of PEC systems employing BIJ and SEJ photoelectrodes can be evaluated under both ideal and practical conditions. Although BIJ and SEJ photoelectrodes exhibit different J–V relationships, the model parameters can be categorized into three consistent, physically meaningful groups: semiconductor parameters, interface parameters, and counter electrode parameters, as summarized in Table 1.

**Table 1. Summary and classification of model parameters.**

| | Semiconductor parameters | Interface parameters | Counter electrode parameters |
|---|---|---|---|
| BIJ | $\beta$, $\theta$, $E_g$, $R_s$ | $J_{ex}$, $R_{sh}$ | $\Delta U$, $J_{ex,c}$ |
| SEJ | $\beta$, $\theta$, $E_g$, $R_s$ | $J_{ex,h}$, $J_{ex,e}$, $R_{sh}$ | $\Delta U$, $J_{ex,c}$ |

To elucidate the role of these parameters on a device J-V curve, two representative cases are considered. For BIJ PEC systems, a photoelectrode with a bandgap of 1.12 eV is used, reflecting the common case of a silicon solar cell photoanode (Fig. 1(d)). For SEJ PEC systems, a photoelectrode with a bandgap of 2.1 eV is instead taken as an example, representing the commonly studied hematite photoanodes (Fig. 1(e)).

In both cases, the shape of the J–V curve is primarily determined by the semiconductor parameters. The parameter $\beta$ reflects carrier recombination, reducing $V_{OC}$ mainly through its effect on the dark current. The parameter $\theta$ governs light absorption and photocarrier generation, and thus is reflected in a reduction in $J_{SC}$. The series resistance $R_s$ introduces additional voltage losses and overall reduces the FF. All these results are in agreement with well-established solar-cell models and collectively, these parameters define the upper performance limit of the photoelectrode through carrier generation, recombination, and transport losses. In contrast, interface parameters primarily affect charge extraction. For the BIJ system (Fig. 1(d)), a decrease in the interface $J_{ex}$ leads to kinetic limitations in charge transfer, resulting in a further reduction in FF. Meanwhile, a decrease in $R_{sh}$ introduces parasitic current pathways, suppressing both $J_{SC}$ and $V_{OC}$. In SEJ case, instead, the interface exchange current density is decomposed into $J_{ex,h}$, and $J_{ex,e}$, corresponding to hole and electron transfer, respectively, with distinct impact on the device response. A decrease in $J_{ex,h}$ leads to a pronounced reduction in FF and evident kinetic limitations near $V_{OC}$, consistent with interface limitations observed in BIJ PEC systems. In contrast, an increase in $J_{ex,e}$ results in a significant reduction in $V_{OC}$, as interface reactions under photoanode operation are hole-dominated, while electron-mediated processes correspond to reverse or parasitic pathways, enhancing recombination and lowering the $V_{OC}$. The impact of the counter electrode on overall PEC performance is further examined in Supplementary Information Section 4, showing that when $J_{ex,c}$ deviates from ideal behavior, interface kinetics at the counter electrode become limiting, leading to additional shifts in the operating point and further power loss.

Overall, the performance of PEC systems is governed by the interplay of semiconductor properties, interface charge transfer, and counter electrode reactions. Supplementary Information Section 4 further quantitatively elucidates the effects of these parameters on the efficiencies of the photoelectrode and the overall PEC system: among them, only the semiconductor bandgap $E_g$ and the reaction potential difference $\Delta U$ exhibit non-monotonic effects, and these two parameters are characteristic of the photoelectrode semiconductor material and the electrolyte reaction system, respectively. The remaining parameters show predominantly monotonic influences within the considered range, but their relative importance to performance varies. These results provide a physical basis for improving PEC performance through material selection, interface matching, and device optimization.

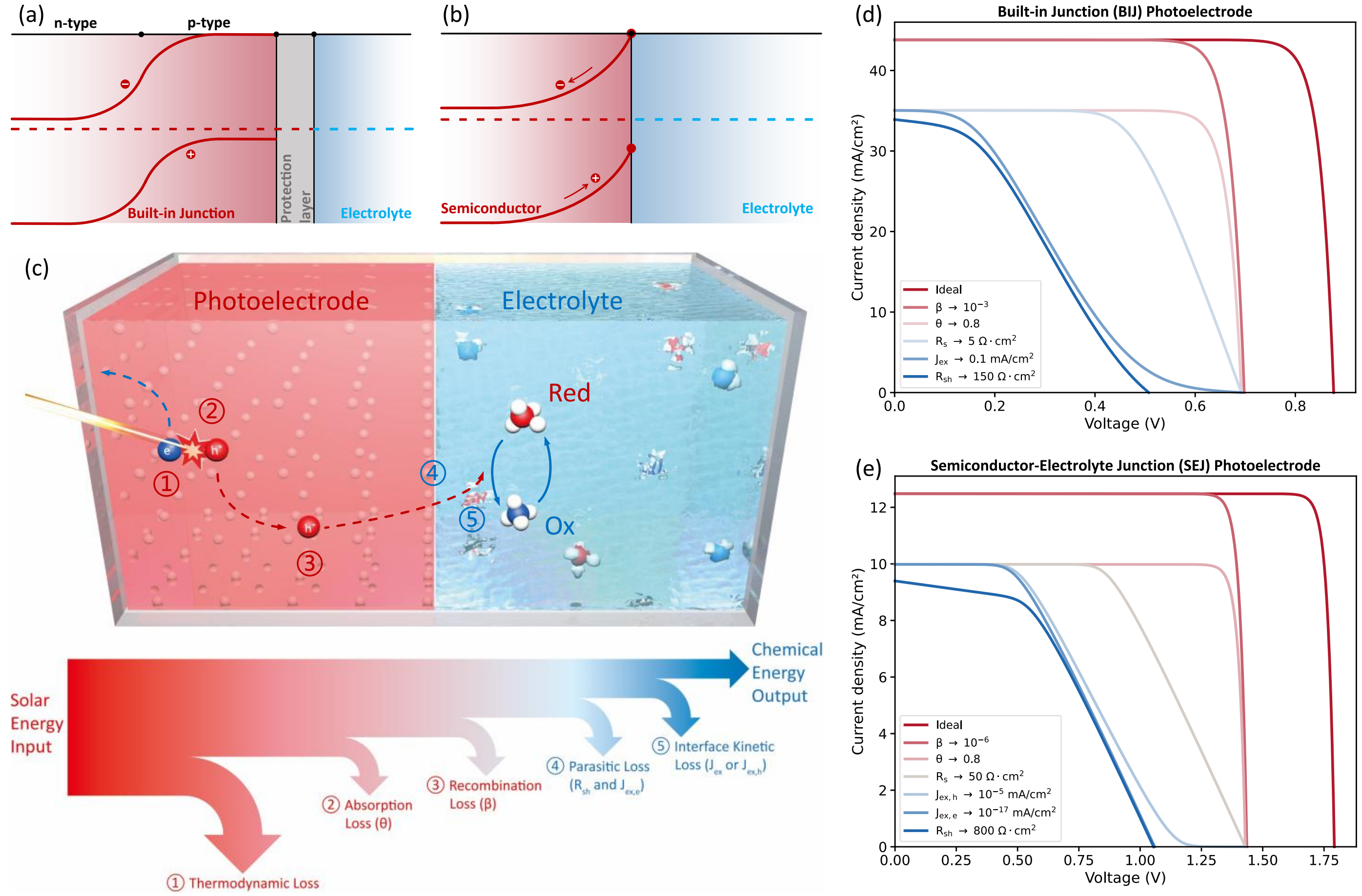

**Figure 1. Schematic illustration of photoelectrode configurations, energy losses, and parameter-dependent J–V characteristics.**

(a) Schematic illustration of a built-in junction (BIJ) photoelectrode;

(b) Schematic illustration of a semiconductor–electrolyte junction (SEJ) photoelectrode.

(c) Schematic illustration of energy loss pathways in photoelectrodes (For energy loss pathways 4 and 5, the second parameter shown in parentheses corresponds to the case of the SEJ photoanode).

(d) J-V curves of a BIJ photoelectrode under variation of model parameters. (Semiconductor bandgap: 1.12 eV. Ideal photoelectrode parameters: $\beta=1$, $\theta=1$, $J_{ex}=\infty$, $R_s=0$, and $R_{sh}=\infty$. Each curve is obtained by changing one parameter relative to the previous curve.)

(e) J-V curves of a SEJ photoelectrode under variation of model parameters. (Semiconductor bandgap: 2.1 eV. Ideal photoelectrode parameters: $\beta=1$, $\theta=1$, $J_{ex,h}=\infty$, $J_{ex,e}=0$, $R_s=0$, and $R_{sh}=\infty$. Each curve is obtained by changing one parameter relative to the previous curve.)

## 2.2. Performance Limits and Efficiency Maps of Ideal and Practical Devices

Based on the obtained J-V relationships, PEC efficiency maps (Fig. 2(a-c)) are constructed as a function of the semiconductor bandgap $E_g$ and the reaction potential difference $\Delta U$ to identify the optimal efficiency regions, providing guidance for material selection and system matching. Under ideal conditions, BIJ and SEJ are equivalent systems and the resulting efficiency map is shown in Fig. 2(a). The upper performance limit for a single junction (solid-solid or solid-liquid) system exceeds 33% and occurs at a bandgap of 1.335 eV and a reaction potential of 0.983 V. However, practical devices deviate significantly from ideal conditions, and the impact of non-ideal loss mechanisms differ between BIJ and SEJ PEC systems. In order to assess practical efficiency limits, the remaining model parameters must first be specified. To do so, our analytical model is utilized to fit the experimental J-V curves of state-of-the-art BIJ devices based on Si, CZTS, CIGS, and emerging materials such as perovskites and $Sb_2Se_3$. Each model parameter is then assigned the best value obtained among all these different studies. The calculated BIJ PEC efficiency map, shown in Fig. 2(b), thus represents the highest achievable performance under current experimental conditions. Similarly, for SEJ PEC systems, representative high-performance studies of $Cu_2O$, $Ta_3N_5$, and $BiVO_4$ are analyzed, and the corresponding model parameters are obtained through curve fitting, followed by selection of optimal parameter combinations. The resulting SEJ PEC efficiency map is shown in Fig. 2(c). Details of the study selection, parameter fitting procedures, and the resulting parameter values are provided in Supplementary Information Section 6.

The maximum PEC efficiency of realistic BIJ PEC systems occurs at a bandgap of 1.386 eV and a reaction potential difference of 0.891 V, with a peak efficiency of 27.16%. In contrast, for SEJ PEC systems, the optimal realistic efficiency is achieved at a bandgap of 2.359 eV and a reaction potential difference of 0.751 V, with a maximum efficiency of 5.63%. The darker red regions correspond to parameter ranges that nonetheless yield high efficiencies. For BIJ PEC systems, the semiconductor bandgap should lie within 1.1-1.6 eV, and the reaction potential difference within 0.6-1.3 V. For SEJ PEC systems, the optimal bandgap range shifts to higher values of approximately 2.0-2.6 eV, while the acceptable range of reaction potentials is broader. These results indicate fundamental differences in material selection strategies between the two systems, both requiring appropriate matching between semiconductor properties and electrolyte reaction characteristics.

In terms of efficiency limits, BIJ PEC systems exhibit higher performance potential under current technological conditions, with optimal bandgaps consistent with those of high-efficiency photovoltaic materials. However, for reactions with large potential differences (e.g., water splitting), commonly used materials deviate from the optimal efficiency region. For example, when the reaction potential difference is fixed at 1.23 V, the realistic maximum efficiency is approximately 24%, and semiconductors with bandgaps larger than 1.7 eV would be required to realize it. When activation overpotentials are further considered, the maximum efficiency decreases to approximately 15%, and the required bandgap increases to above 2 eV. In contrast, SEJ PEC systems exhibit a wider tolerance to reaction potential differences, with most practical reactions falling within acceptable efficiency regions, making them more suitable for high-potential reactions. However, as shown by the listed model parameters for this case, large activation overpotentials still lead to significant efficiency losses.

Furthermore, efficiency maps necessarily neglect the relative alignment between a specific photoelectrode material and electrolyte Fermi level. In SEJ PEC systems, however, the electric field required for carrier separation solely originates from this energy difference and the associated band-bending effect. To understand this interaction, the theoretical expressions of $J_{ex,h}$ and $J_{ex,e}$ (Eqs. S48 and S49) need to be considered, which depend on the semiconductor band-edge energies ($E_V$, $E_C$), the electrolyte Fermi level ($E^{el}$) and the reactant reorganization energy $\lambda_r$. For a representative n-type semiconductor (Fig. 2(d1)), the hole (electron) exchange current $J_{ex,h}$ ($J_{ex,e}$) increases when the electrolyte energy level $E^{el}$ is closer to $E_V$ ($E_C$). Notably, the influence of the reorganization energy $\lambda_r$ on $J_{ex,h}$ and $J_{ex,e}$ is nonmonotonic within the semiconductor bandgap range, with $J_{ex,h}$ increasing (decreasing) with $\lambda_r$ for $E^{el}$ approaching $E_V$ ($E_C$). Also, as shown in Fig. 2(d2), both $J_{ex,h}$ and $J_{ex,e}$ exhibit extrema as $\lambda_r$ varies. For n-type semiconductors, the peak value of $J_{ex,h}$ becomes larger for $E^{el}$ closer to the valence band and smaller $\lambda_r$ values. Consistent conclusions are also obtained for p-type semiconductors (see Supplementary Information Section 2). These results indicate that, in SEJ PEC systems, the matching between the photoelectrode and the electrolyte is not a simple problem of energy-level alignment, but rather a cooperative optimization process jointly determined by energy-level positions, interface reaction kinetics, and the concentrations and physicochemical properties of redox species. Together, these results provide clear design guidance for the selection of photoelectrode and electrolyte systems.

## 2.3. Quantitative Loss Analysis of Ideal and Practical Devices: From Diagnostic to Optimization

Using the model parameters and the optimal bandgaps identified from the BIJ and SEJ efficiency maps presented in Section 2.2, a Shapley allocation method[35] (see Supplementary Information Section 5) is implemented to quantitatively evaluate the energy loss contributions associated with each model parameter, enabling the construction of Sankey diagrams of energy flow for both systems (Fig. 2(e,f)). These diagrams visually illustrate the energy flow of the input 100 mW/cm$^2$ AM1.5G solar energy within the photoelectrode-electrolyte system.

For BIJ PEC system, due to the use of smaller bandgap semiconductors, although the thermodynamic loss corresponding to the Shockley-Queisser limit remains dominant, its proportion is significantly lower than that in SEJ photoelectrode. This loss is primarily determined by the bandgap and is difficult to further reduce once the material is fixed; therefore, it is not the focus of this work. In contrast, the Sankey diagram clearly shows that the major energy losses in BIJ devices originate from semiconductor parameters (5.9%), while losses associated with interface parameters are nearly negligible (0.5%). This indicates that current high-performance BIJ photoelectrodes have achieved near-ideal interface charge transfer. Therefore, further performance improvement in BIJ systems mainly requires reducing semiconductor losses. Among these, the recombination loss coefficient $\beta$ and the series resistance $R_s$ contribute most significantly. The former is associated with non-radiative recombination, indicating that recombination losses remain a primary limiting factor, while the latter reflects transport-related losses arising from low carrier mobility and high resistive components, including the conductivity of different layers within the photoelectrode as well as the overall PEC device design. Accordingly, future optimization should focus on suppressing recombination and improving charge transport in the device.

For SEJ photoelectrode systems, in contrast to BIJ systems, the energy losses arising from interface parameters (5.9%) are greater than those from semiconductor parameters (3.7%). Regarding interface losses, the exchange current density associated with minority carriers (corresponding to $J_{ex,h}$ for photoanodes and $J_{ex,e}$ for photocathodes) contributes most significantly to the total loss, reflecting insufficient interface reaction kinetics. This is consistent with the fact that high-performance SEJ photoelectrodes are mainly applied to solar water splitting, which inherently involves sluggish reaction kinetics. For semiconductor losses, the energy losses associated with series resistance are comparable to the recombination-related losses. This behavior may be attributed to the generally low carrier mobility and the high density of defect states in medium-bandgap semiconductors.

A wide range of strategies has been developed over the decades to improve the performance of semiconductor materials and photoelectrodes, such as heterojunction or doping and nanostructuring or co-catalyst deposition. Table 2 summarizes commonly used approaches for improving photoelectrode performance and establishes a mapping between these strategies and model parameters according to their effects. The direct link between performance enhancement strategies and key model parameters allows the results of the above energy loss analysis to be directly translated into specific directions for material and device optimization methods, thereby providing guidance for the design and performance improvement of photoelectrodes. Specifically, for BIJ photoelectrodes, strategies that reduce semiconductor recombination should be prioritized; for SEJ photoelectrodes, it is necessary to enhance interface reaction kinetics.

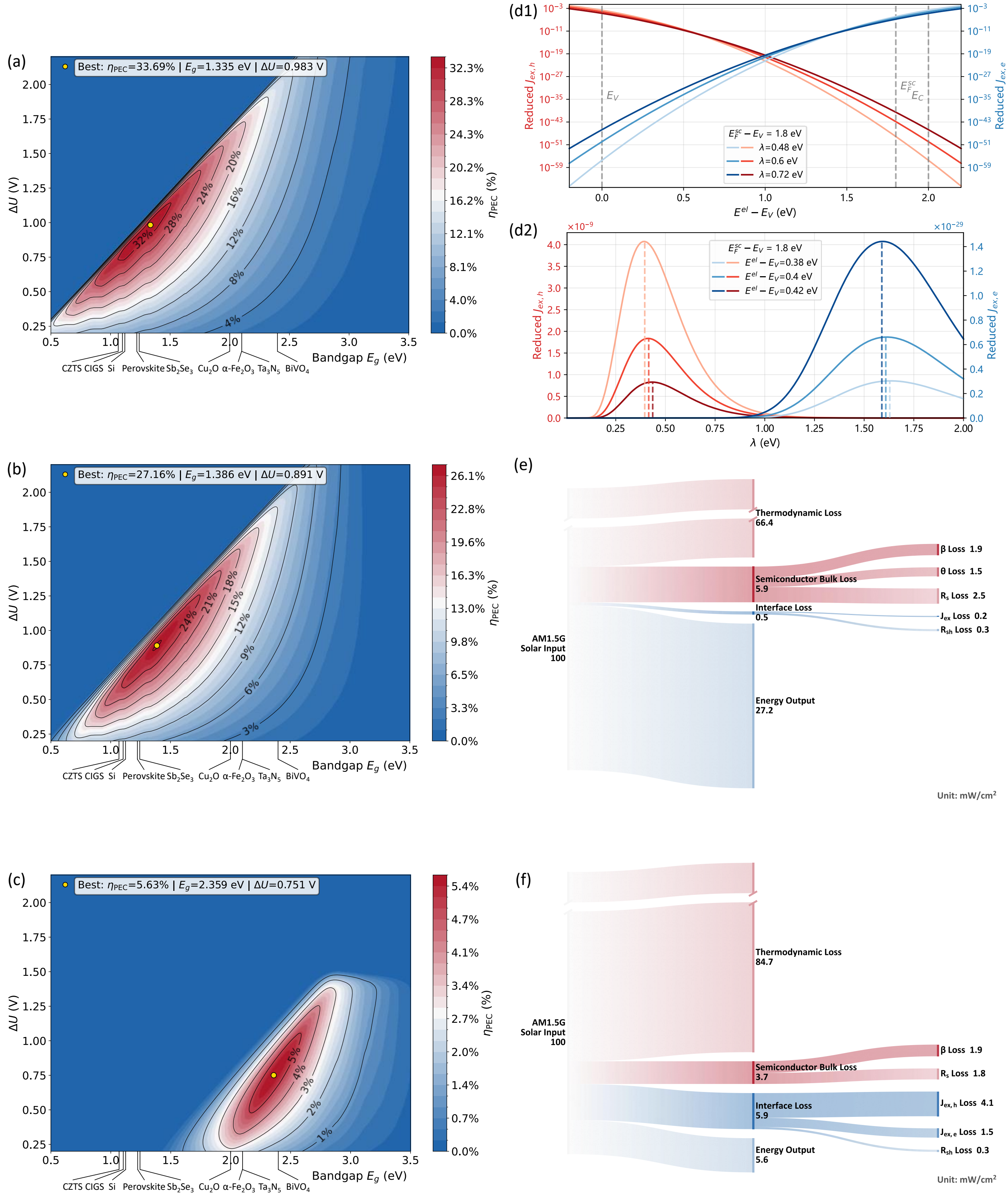

**Figure 2. Efficiency limits, efficiency maps, semiconductor–electrolyte matching, and energy-loss analysis of photoelectrochemical cells.**

(a) Efficiency map under ideal conditions (same for BIJ and SEJ, using the ideal photoelectrode parameters defined in Fig. 1.);

(b) Efficiency map for the BIJ PEC cell; (β= $1.000\times10^{-1}$, θ= 0.9515, $J_{ex}$= $1.399\times10^{2}$ mA/cm², $R_s$= 2.540 Ω cm², and $R_{sh}$= $2.671\times10^{3}$ Ω cm², details are provided in Supplementary Information Section 6);

(c) Efficiency map for the SEJ PEC cell; (β= $8.066\times10^{-6}$, θ= 1.000, $J_{ex,h}$= $3.515\times10^{-11}$ mA/cm², $J_{ex,e}$= $1.005\times10^{-25}$ mA/cm², $R_s$= 30.05 Ω cm², and $R_{sh}$= $5.166\times10^{3}$ Ω cm², details are provided in Supplementary Information Section 6);

(In the efficiency maps, the counter electrode is assumed to be ideal, with the exchange current density $J_{ex,c}$=∞.)

(For tunable-bandgap semiconductors such as perovskites, CIGS, and CZTS, the bandgap values shown in the figure correspond to the specific materials reported in the respective references. Details of the literature selection are provided in Supplementary Information Section 6.)

(d1) Dependence of interface exchange current densities on electrolyte energy-level position;

(d2) Dependence of interface exchange current densities on reorganization energy;

(e) Energy-loss Sankey diagram for the BIJ photoelectrode at the optimal bandgap identified from the BIJ efficiency map, with model parameters consistent with those used in Fig. 2(b);

(f) Energy-loss Sankey diagram for the SEJ photoelectrode at the optimal bandgap identified from the SEJ efficiency map, with model parameters consistent with those used in Fig. 2(c) (The parameter associated with interface loss is shown for the photoanode as an example).

**Table 2. Mapping between performance improvement strategies for photoelectrodes and model parameters.**

| Improvement strategies | Semiconductor parameters | | | | Semiconductor-electrolyte interface parameters | |
|---|---|---|---|---|---|---|
| | $E_g$ | β | θ | $R_s$ | $J_{ex}$ [(1)] | $R_{sh}$ |
| Nanostructure | ○ | √ | √ | √ | √ | ○ |
| Heterojunction | | √ | | × | ○ | √ |
| Doping | ○ | √ | | √ | | ○ |
| Co-catalyst | | √ | | ○ | √ | |
| Back contact | | √ | | √ | | |
| Surface passivation | | √ | | × | ○ | √ |
| Facet engineering | | √ | | | √ | √ |
| Surface texturing | | | √ | | √ | ○ |
| Surface plasmon resonance | | | √ | | √ | |

Note: The symbols used in the table are defined as follows. A check mark (√) indicates that the corresponding strategy generally improves the parameter; a cross (×) indicates a potential negative impact; a circle (○) indicates that the effect depends on the specific system and is therefore uncertain; a blank entry indicates that the influence is typically small or negligible. (1): $J_{ex}$ denotes the interface exchange current density; for BIJ systems, it represents the interface exchange current density, while for SEJ systems, it corresponds to the minority-carrier exchange current density (e.g., $J_{ex,h}$ for photoanodes).

# 3. Model Validation and Analysis of PEC Materials and Devices

Based on the design principles established for PEC systems, this section applies the framework to specific material systems. A systematic comparison is conducted among different materials in terms of their theoretical efficiency limits, practically achievable performance, dominant limiting mechanisms, and application potential. On this basis, optimization strategies for material selection and system design are proposed.

## 3.1. Theoretical Limits and Practical Performance Upper Bounds of Materials

Semiconductor materials currently employed in photoelectrodes can be broadly categorized into two groups. One group consists of photovoltaic (PV) grade materials that have already been applied in solar cells, such as Si, CIGS, and CZTS, as well as emerging materials such as perovskites and $Sb_2Se_3$, while the other includes non-photovoltaic grade (non-PV grade) materials that have primarily been developed within PEC systems, primarily oxide semiconductors ($\alpha$-$Fe_2O_3$, $Cu_2O$, and $BiVO_4$) and nitride semiconductors ($Ta_3N_5$)[i].

[i] In addition, commonly studied non-PV grade semiconductors include $TiO_2$, $WO_3$, InGaN, and two-dimensional materials such as g-$C_3N_4$ and $MoS_2$. Among them, $TiO_2$, $WO_3$ can serve as standalone absorber layers; however, their large bandgaps limit their applicability in high-efficiency PEC systems, and thus they are not further discussed in this work. The remaining materials are mostly used as auxiliary components to enhance performance rather than as primary absorber layers, and are therefore also excluded from the present analysis.

Fig. 2(b-c) map the above materials onto the efficiency maps of BIJ and SEJ PEC systems, respectively. PV grade semiconductors generally exhibit higher performance potential in BIJ PEC systems, whereas the bandgap range of non-PV grade semiconductors is closer to the optimal region in the SEJ PEC efficiency map. Material matching becomes particularly critical when considering reaction constraints. For water splitting, corresponding to a thermodynamic potential of 1.23 V (and higher when overpotentials are considered), the set of viable materials in both BIJ and SEJ systems is largely restricted to non-PV semiconductors.

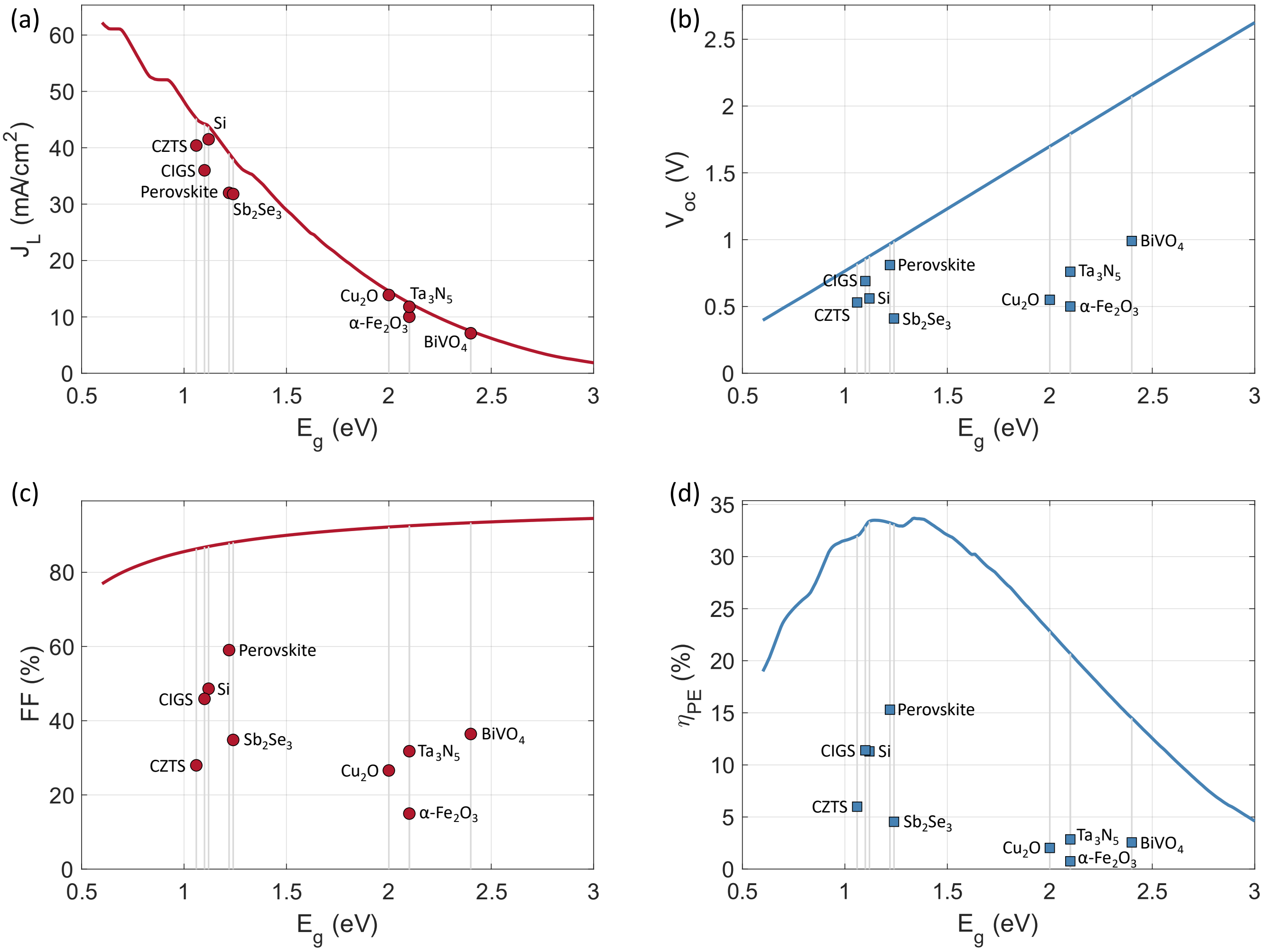

**Figure 3. Theoretical limits and state-of-the-art performance of photoelectrochemical cell materials.**
(a) Maximum photocurrent density vs. bandgap;
(b) Open-circuit voltage vs. bandgap;
(c) Fill factor vs. bandgap;
(d) $\eta_{PE}$ vs. bandgap.
(For tunable-bandgap semiconductors such as perovskites, CIGS, and CZTS, the bandgap values shown in the figure correspond to the specific materials reported in the respective references. Details of the literature selection are provided in Supplementary Information Section 6.)

Based on Eq. 7, the theoretical limits under ideal conditions of photocurrent density, open-circuit voltage, fill factor, and photoelectrode efficiency can be consistently derived for both BIJ and SEJ configurations, as both reduce to the diode equation. These limits, together with experimentally reported state-of-the-art results up to 2025 selected based on outstanding photocurrent density, are summarized in Fig. 3(a-d), enabling a direct comparison between theoretical performance bounds and practical realizations. While this analysis is routine for photovoltaic devices, to the best of our knowledge it is the first time all four figures of merit are compared against their theoretical limits for PEC photoelectrodes.

This comparison reveals a clear divergence: although the photocurrent density of most materials approaches their theoretical limits, the $V_{OC}$ and FF remain significantly below their ideal values, particularly for non-PV semiconductors.

This mismatch partly arises from a bias in evaluation criteria. Indeed, high-performance materials are typically assessed using the maximum photocurrent density as the primary metric. Photoelectrode efficiency, by contrast, requires evaluation of the maximum power point, which varies between materials and architectures and is thus harder to compare. Accordingly, we also adopted this as the key selection metric for studies to be included in the comparison.
From a mechanistic perspective, in contrast to $J_{SC}$, the $V_{OC}$ is highly sensitive to non-radiative recombination processes, i.e. $\beta$, as discussed in Section 2. As a result, device performance may still be limited even under conditions of near-ideal photocurrent. Meanwhile, sluggish interface reaction kinetics represents another key bottleneck in PEC systems, introducing additional charge transfer barriers and thereby significantly reducing the FF. Experimentally, enhancing light absorption through material design improves photocurrent density but does not directly address suppression of non-radiative recombination and optimization of interface reaction kinetics. This explains why improvements in photocurrent do not necessarily translate into corresponding enhancements in $V_{OC}$, FF, and the overall energy conversion efficiency.
Taken together, these results quantify an underlying design trade-off in PEC systems: existing PV grade semiconductors, with their smaller bandgaps and better intrinsic properties, are capable of approaching efficiency limits in BIJ configurations but are fundamentally constrained in high-potential reactions, whereas non-PV semiconductors, with larger bandgaps, are better suited for such reactions but suffer from lower overall efficiency. Importantly, optimal performance across both material classes is consistently achieved within a moderate reaction potential window of approximately 0.7-1.1 V, defining a universal operating regime for high-efficiency PEC systems.

## 3.2. Bottleneck Analysis and Improvement Directions of Materials

Based on the energy loss analysis framework established above, the experimental J-V curves of the chosen representative, state-of-the-art PEC studies are fitted using Eqs. 5 and 6, enabling quantitative energy-loss decomposition to identify performance bottlenecks and corresponding optimization strategies. Specifically, five PV grade and four non-PV grade based photoelectrodes were included in the comparison. In all cases, high fitting quality is achieved consistently, validating the general applicability of the model. The corresponding material parameters, fitted curves, and $R^2$ values are provided in Supplementary Information Section 6.
For Si photoelectrode, Fan et al. developed an $n^+np^+$-Si BIJ architecture for the hydrogen evolution reaction (HER), achieving a photocurrent density of 41.5 mA/cm$^2$ by integrating a Ni protective/electron transport layer and $MoS_2/Ni_3S_2$ catalysts[36]. Fig. 4(a) visualizes the distribution of energy-loss contributions for the reference $n^+np^+$-Si +Ni sample and the improved $n^+np^+$-Si +Ni+ $MoS_2/Ni_3S_2$ one. It quantitatively shows that the introduction of the catalyst layer dramatically reduces the interfacial losses from 7.2% to sub 1%, thanks to enhanced interface charge-transfer kinetics ($J_{ex}$ increases). Concurrently, semiconductor losses are also reduced, with recombination and shunt resistance losses each decreasing by 1%. These analytical results are in good agreement with experimental expectations. This analysis however reveals that additional improvements of this device architecture will require a reduction in $\beta$-related losses as interfacial effects are no longer significant. Table 2 thus helps to identify viable strategies for bringing the performance of this Si photoelectrode closer to its theoretical limit.
For the $Ta_3N_5$ photoelectrode, Zhao et al.[37] fabricated a $Ta_3N_5$-$AlO_x$/Fh-$NiFeO_x$/$CeO_x$ structured SEJ architecture for the oxygen evolution reaction (OER), achieving a photocurrent density of 11.8 mA/cm$^2$. In this system, AlOx serves as a passivation layer, Fh as a hole transport layer, $NiFeO_x$ as the OER catalyst, and $CeO_x$ enhances catalyst stability. Fig. 4(b) reports the energy-loss decomposition for the reference bare $Ta_3N_5$ and the optimized device. It shows that the reference device already showed good performance, with only interfacial kinetics and series resistance losses each contributing more than 5% performance reduction. It also highlights that the introduction of passivation and hole transport layers results in a performance gain via a reduction in semiconductor losses (from 9% to 6.8%), accompanied by improvements in $\beta$ and $R_s$ corresponding to reduced recombination and better carrier transport. The catalyst layer results in a 0.7% decrease in interface losses, dominated by the increase in exchange hole current density $J_{ex,h}$ reflecting markedly enhanced interface hole reaction kinetics. These quantitative model-based analyses are also highly consistent with experimental observations. Contrary to the above Si photoelectrode, the primary loss channel for the optimized $Ta_3N_5$ remains the limited exchange hole current (6.9% losses), and thus a better catalyst selection could further improve its performance. Concurrently, among the bulk semiconductor losses, $R_s$ still contributes to 4.3% performance reduction and would be the second aspect to improve.

The same methodology is applied to representative studies of the remaining seven materials and Fig. 4(c) summarizes their energy loss distributions. Overall, PV grade materials exhibit higher efficiencies and lower thermodynamic losses compared to non-PV grade materials. For PV-grade materials, energy losses are primarily dominated by semiconductor processes, among which losses associated with $R_s$ and $\beta$ account for the major contributions. Therefore, further optimization of PV materials is expected to be mainly achieved by suppressing bulk recombination within the semiconductor and improving charge transport. At the same time, for certain materials, interface losses are also non-negligible, which mainly originate from limitations in interface reaction kinetics, indicating that optimization of interface processes is also of importance. In contrast, for non-PV grade materials, energy losses are primarily governed by interface processes, with reaction kinetics being the dominant limiting factor; therefore, optimizing interface reaction kinetics is expected to deliver more substantial performance improvements. Furthermore, as shown in Fig. 4(c), the current performance-limiting bottlenecks of specific materials can be directly identified. These bottlenecks can be further correlated with the strategy-parameter mapping summarized in Table 2, thereby providing direct guidance for material selection and device design.

It is also worth noting that all nine optimal devices discussed above are applied to solar water splitting systems. To further demonstrate the broad applicability of the model, this approach is extended to reported solar $CO_2$ reduction, $NH_3$ synthesis, and solar redox flow battery systems. In all cases, the model yields consistently high-quality fits with strong agreement between model predictions and experimental results, underscoring its robust generality across PEC solar energy conversion and storage systems (see Supplementary Information Section 7).

## 3.3. Comparison of Material Systems and Future Design Strategies

Although energy loss analysis can effectively identify performance bottlenecks in material systems, a single performance metric is insufficient to support long-term assessments of material development. Therefore, a multidimensional evaluation framework is finally introduced to systematically compare photoelectrode materials in terms of resource sustainability, device stability, performance limits, and environmental impact, thereby addressing the key question of which material systems hold the greatest future potential.

Specifically, five evaluation dimensions are considered: elemental abundance, operational stability, state-of-the-art efficiency, theoretical efficiency, and toxicity. Elemental abundance is quantified using the crustal abundance of the scarcest element in the material composition, reflecting material cost and scalability. Stability is defined as the operating time corresponding to a 1% decay in photocurrent, characterizing practical operational reliability. The efficiency metrics correspond to the current state-of-the-art (using the maximum photocurrent density as the evaluation metric, as discussed above) experimental performance and the theoretical limit, respectively. Toxicity is evaluated based on the acute toxicity classification of the materials and their precursors (detailed evaluation methods and literature selection are provided in Supplementary Information Section 8). The corresponding results are presented as radar plots in Fig. 4(d,e).

Among PV grade materials, Si exhibits the most balanced overall performance, characterized by its high abundance, low toxicity, and excellent stability and efficiency enabled by mature processing technologies. In contrast, perovskite materials, although achieving high levels of efficiency and stability, remain potentially limited by toxicity and elemental scarcity. Materials such as CIGS, CZTS, and $Sb_2Se_3$ are currently constrained by stability and practical efficiency, all exhibiting significant room for improvement. Among them, $Sb_2Se_3$, owing to its comparatively high theoretical efficiency, may play an important role in future sustainable PEC systems.

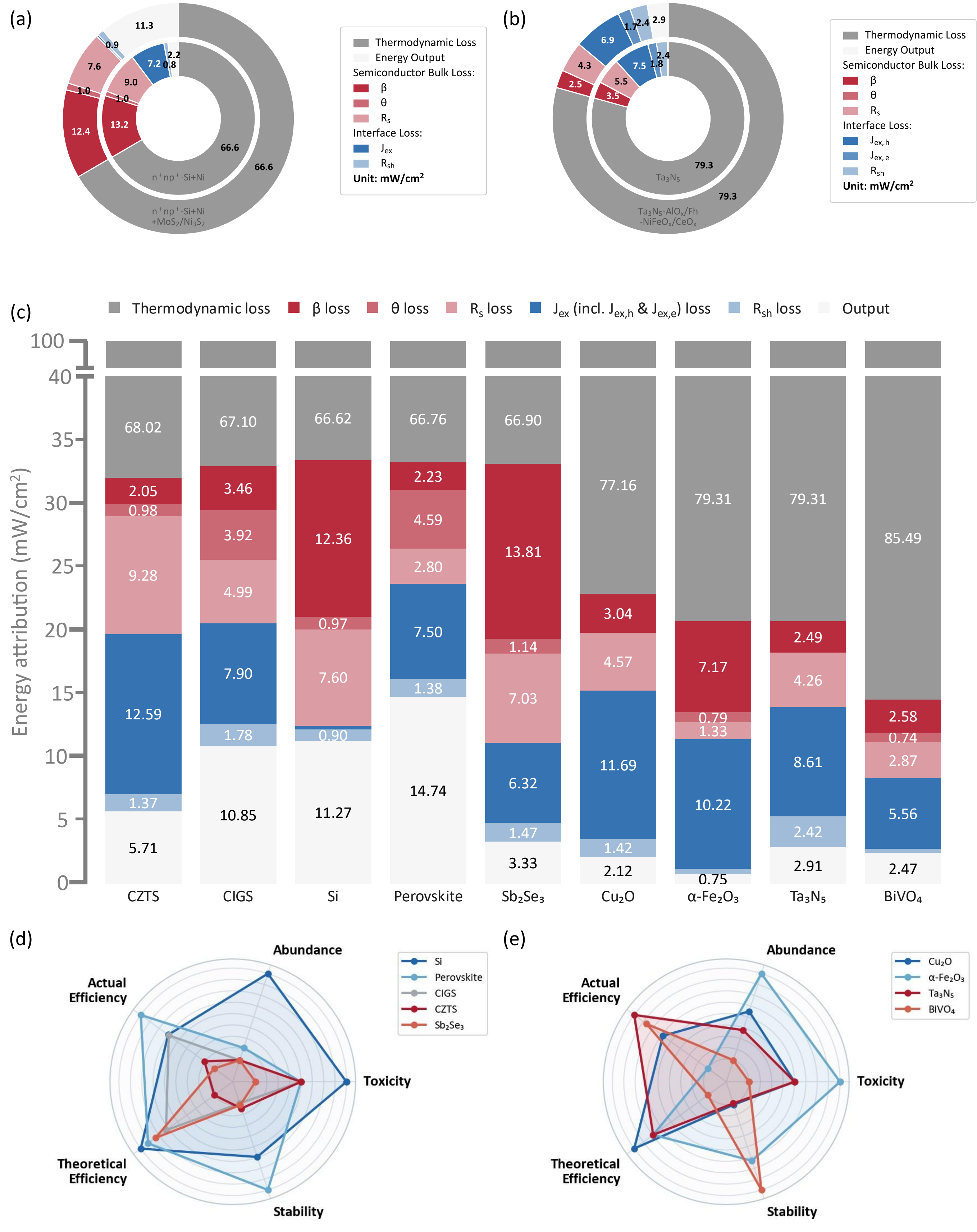

**Figure 4. Energy loss analysis and comparative evaluation of representative photoelectrodes materials.**

(a) Donut chart of the Si photoelectrode;

(b) Donut chart of the $Ta_3N_5$ photoelectrode;

(c) Energy loss distribution of nine representative state-of-the-art materials;

(d) Radar chart for comprehensive comparison of PV grade materials;

(e) Radar chart for comprehensive comparison of non-PV grade materials.

For non-PV grade materials, α-$Fe_2O_3$ offers significant advantages in terms of cost and environmental compatibility, yet its practical efficiency remains far below its theoretical potential. $Ta_3N_5$ has achieved relatively high performance to date; however, its stability still requires improvement, and its low elemental abundance further limits its scalability. $BiVO_4$ exhibits excellent operational stability, but its efficiency ceiling is constrained by intrinsic material properties. In contrast, $Cu_2O$ possesses high theoretical efficiency potential and is composed of earth-abundant elements, but suffers from severe stability issues. Overall, no non-PV material currently demonstrates simultaneous superiority across all evaluation dimensions, indicating that substantial breakthroughs are still required.

Cross-system comparisons further reveal that PV grade materials generally outperform non-PV grade materials in terms of efficiency and stability. This advantage stems from the technological maturity of photovoltaics and their superior intrinsic material properties. However, non-PV materials typically possess larger bandgaps and benefit from simpler electrode architectures, resulting in lower costs. These features render them indispensable in specific application scenarios, such as high-potential reactions or low-cost large-scale deployment.

In terms of material selection, PV grade materials have traditionally been used to construct BIJ configurations, whereas non-PV materials are primarily employed in SEJ systems. However, this empirical classification lacks fundamental physical necessity. Future research should, on the one hand, explore cross-paradigm architectures, such as employing non-PV materials in high-performance BIJ photoelectrode or incorporating PV materials into SEJ systems, as well as developing multijunction or hybrid structures. On the other hand, efforts should focus on developing PV grade materials with larger bandgaps (>1.6 eV) to expand their applicability in high reaction-potential systems.

Regarding performance improvement strategies, beyond conventional defect passivation and interface engineering, emerging low-dimensional materials, such as two-dimensional materials and quantum dots, provide new degrees of freedom for tuning carrier dynamics. These materials are expected to play a critical role in suppressing recombination and enhancing interface reaction kinetics.

With respect to reaction system selection, high-efficiency PEC systems are currently dominated by solar water splitting. Reactions such as $CO_2$ reduction have also been extensively studied, yet their performance still lags significantly behind that of water splitting systems. Therefore, future research should focus on expanding reaction types to better match photoelectrode properties and improving the performance of non-water-splitting systems, including $CO_2$ reduction, high-value chemical synthesis, and solar redox flow batteries based on fast and reversible redox couples.

At the theoretical level, although existing models can reasonably describe energy loss distributions in PEC systems, a more detailed understanding of the coupling between semiconductor charge behavior and interface charge transfer in SEJ systems is still lacking. Solar redox flow batteries involve fast and reversible redox reactions, leading to weaker interface kinetic limitations and thereby reducing the dominance of interface processes over overall performance. This characteristic makes them an ideal model system for studying carrier generation, recombination, and transport processes. Therefore, solar redox flow batteries can serve as an ideal platform for further developing and refining theoretical frameworks capable of accurately describing bulk–interface coupling, thereby systematically elucidating the fundamental mechanisms governing PEC systems.

# 4. Strengths and Limitations of the Model

The strengths and limitations of theoretical models can usually be evaluated from the following four dimensions: (i) assumptions: the validity of the underlying assumptions and their ability to capture the essential physical and chemical processes without oversimplification or neglect of critical mechanisms; (ii) parameters: the proper definition of model parameters, including their number, physical interpretability, degree of lumping and coupling, and whether they are measurable and uniquely identifiable; (iii) applicability: the consistency of the model with experimental observations, as well as clearly defined conditions and ranges of validity; (iv) predictive capability: the ability of the model to provide reliable predictions beyond existing data, particularly in guiding experimental design and system optimization.

**Assumptions:** This model is based on a series of simplifying assumptions to enable an analytical formulation. For interface reactions, the electrode-electrolyte interface is assumed to follow a single-electron, single-step reaction process, and the involvement of surface states and the resulting Fermi level pinning effect are neglected, thereby describing interface charge transfer as an idealized process. In the semiconductor region, the low-injection approximation is adopted, and recombination within the space-charge region is assumed to be absent, thereby simplifying the solution of carrier transport and distribution. The model further assumes a one-dimensional structure, neglecting spatial inhomogeneity and the effects of complex geometries, and does not consider mass transport limitations in the electrolyte. Furthermore, the model does not consider the introduction of a catalyst layer on the semiconductor surface that prevents direct contact between the semiconductor and the electrolyte, and therefore cannot describe changes in the charge transfer mechanism under such configurations; additionally, the model is established based on a single-junction structure and does not include the coupling between different junctions or carrier transport behavior in multi-junction systems.

Overall, these assumptions allow the model to capture the main physical processes while maintaining simplicity and analytical tractability. However, its applicability may be limited in systems where significant surface states, strong recombination, mass transport limitations, or complex interfacial reaction pathways are present.

**Parameters:** The parameters in the model are designed to balance physical interpretability and simplicity, although a certain degree of lumping and coupling is unavoidable. Specifically, the parameter $\beta$ reflects multiple non-radiative recombination processes in the system; its physical origin is relatively complex, but it can be indirectly determined through experimental measurements of carrier lifetime. The parameter $\theta$ is not only related to the light absorption process, but is also influenced by the carrier diffusion length, which affects the conversion of photogenerated carriers into photocurrent; since the diffusion length is closely related to carrier lifetime, $\theta$ thus embodies the coupled effects of multiple physical processes. The interface exchange current density $J_{ex}$ (including both electrons and holes) can be obtained from Tafel plots under dark conditions. However, in practical systems, interface charge transfer is often jointly influenced by interface kinetics, surface states, and other non-ideal factors.

In addition, the series resistance $R_s$ and shunt resistance $R_{sh}$ are introduced as lumped parameters. The physical origin of $R_s$ is usually well defined, such as the bulk resistance of the semiconductor, the photoelectrode substrate, and resistances within the overall device structure, whereas the origin of $R_{sh}$ is relatively complex and is typically associated with non-ideal factors such as interfacial defects and leakage pathways. Nevertheless, these parameters can all be obtained through experimental techniques, such as electrochemical impedance spectroscopy (EIS).

At present, in most related studies, the key parameters involved in such models are often not systematically measured or independently characterized. Future work should therefore focus on systematic experimental measurement and validation of these parameters to further assess the applicability of the model and improve its predictive capability.

Overall, although some parameters represent lumped effects of multiple physical processes and exhibit a certain degree of coupling, their values still maintain clear physical correspondence and can be validated experimentally. Combined with the reasonable fitting strategy implemented in this work, the model can achieve unique parameter determination, thereby ensuring the reliability of parameter extraction.

**Applicability:** The model demonstrates good general applicability, as it can effectively fit most experimentally measured J-V curves, indicating that it provides a reasonable description of the main physical processes. However, in some systems, particularly when the J-V curves exhibit complex behaviors (such as non-ideal inflection points or multi-stage slope variations), the descriptive capability of the model decreases. In addition, for certain samples, the fitting accuracy is also limited (see Supplementary Information, Sections 6 and 7), suggesting that the model does not fully capture all dominant processes in these systems. Therefore, the model is more suitable for systems that satisfy its underlying assumptions, such as those with relatively simple interface reactions, weak mass transport limitations, limited influence of surface states, and relatively homogeneous semiconductor properties.

**Predictive capability:** In terms of predictive capability, the model has strong practical value. By varying model parameters, it can effectively predict how changes in material properties or interface characteristics influence the shape of the J-V curves, thereby providing a quantitative basis for performance optimization. Furthermore, the mapping established between model parameters and physical processes enables the model to guide material selection, electrode-electrolyte matching, and device structure optimization. However, it should be noted that this predictive capability still relies on the validity of the model assumptions and the proper determination of parameters. When operating outside the model's applicable range, its predictions still require further experimental validation.

# 5. Conclusion

This work establishes a unified theoretical model and analysis framework for PEC cells. The framework is applicable to both BIJ PEC cells and SEJ PEC cells, enabling a unified description of the performance limits and loss distributions of these two representative PEC architectures under both ideal and practical conditions. Furthermore, the J-V relationship of semiconductor-electrolyte junction photoelectrodes is systematically derived, explicitly addressing key assumptions and intermediate steps that are often simplified in existing studies, thereby improving the transparency and traceability of the model.

The model is then employed to assess theoretical limits, guide material selection, and trace opportunities for further performance improvement. Efficiency maps are first constructed to elucidate the coupled effects of semiconductor bandgap and reaction potential difference on device performance, thereby identifying the optimal material selection windows and efficiency limits for different PEC systems under ideal and experimentally achievable conditions. The results indicate that BIJ systems exhibit higher theoretical efficiency potential under current conditions, whereas SEJ systems demonstrate greater adaptability to varying reaction potentials. Notably, both material classes achieve optimal performance within a moderate reaction potential window of approximately 0.7–1.1 V, defining a universal operating regime for high-efficiency PEC systems.

Crucially, the model enables quantitative energy loss analysis for different PEC systems and specific materials, allowing key performance bottlenecks to be identified and assigned to three loss channels: thermodynamic losses, semiconductor losses (recombination, light absorption, and series resistance), and interface losses (reaction kinetics and parasitic shunt). The analysis reveals a systematic divergence between the two architectures: in state-of-the-art BIJ photoelectrodes, interface losses have already been reduced to approximately 0.5%, approaching the ideal limit, such that further performance gains require primarily a reduction in recombination (β) and carrier transport losses ($R_s$); in contrast, for SEJ systems, interface losses (5.9%) exceed semiconductor losses (3.7%), making the optimization of interface reaction kinetics the most impactful improvement direction. Concrete device-level examples illustrate this distinction: for the best-performing Si-based BIJ photoelectrode, catalyst integration reduced interface losses from 7.2% to below 1%, demonstrating that further optimization of this device should focus on suppressing bulk recombination rather than interface engineering. Conversely, the dominant loss channel in the best-performing $Ta_3N_5$ SEJ photoelectrode remains the limited hole exchange current density ($J_{ex,h}$, contributing 6.9% loss), identifying improved co-catalyst design as the primary route to further performance gains. Building upon this, a direct correspondence between model parameters, energy loss channels, and performance improvement strategies is established, enabling different loss pathways to be explicitly mapped onto specific material and device optimization routes, thereby translating model insights into actionable design guidelines. The framework is further validated against a broad range of state-of-the-art PEC devices spanning solar water splitting, $CO_2$ reduction, $NH_3$ synthesis, and solar redox flow batteries, confirming its predictive capability and general applicability.

At the interface level, a semiconductor–electrolyte matching criterion that goes beyond simple energy-level alignment is proposed based on the model. This criterion indicates that interface performance is governed by the coupled effects of multiple factors, including band-edge positions, reaction potential, reorganization energy, and redox species concentration, thereby providing a more rigorous physical basis for the co-design of electrolytes and semiconductors in SEJ systems.

Finally, a multidimensional evaluation framework is further constructed, incorporating elemental abundance, operational stability, state-of-the-art performance, theoretical efficiency, and toxicity, to systematically compare representative and emerging photoelectrode materials. The results show that photovoltaic grade semiconductors generally outperform non-photovoltaic grade semiconductors in terms of efficiency and stability, whereas non-PV materials exhibit unique advantages in high reaction-potential systems and low-cost applications. This highlights that material selection in PEC systems is inherently a multi-objective optimization problem, for which no single material system is universally optimal.

In summary, the core contribution of this work lies in establishing a unified model and analysis framework for photoelectrochemical systems, which enables a consistent description, quantitative performance evaluation, and energy loss analysis of different PEC architectures. By guiding material selection, interface matching, and performance optimization, it therefore enables a shift from empirical device optimization toward mechanism-based rational design, which will be key to achieve PEC devices that realize their full potential.

# Supplementary Information for

## *From Loss Diagnosis to Rational Design: A Unified Analytical Model for Photoelectrochemical Cells*

Ziyan Pan, Giulia Tagliabue*
Laboratory of Nanoscience for Energy Technologies (LNET), Institute of Mechanical Engineering, School of Engineering, École Polytechnique Fédérale de Lausanne (EPFL), CH-1015 Lausanne, Switzerland
*Corresponding author: giulia.tagliabue@epfl.ch

# 1. Theoretical Derivation of Photoelectrode Current Density–Voltage Relationships

## 1.1. Built-in Junction Photoelectrodes

Taking a photoanode as an example, as illustrated in Fig. S1(a), the photoelectrode consists of a p-n junction, a protective layer, and an electrolyte in contact with the electrode surface. Under dark equilibrium conditions, the Fermi levels of the semiconductor, protective layer, and electrolyte are aligned. Under illumination, the electron and hole quasi-Fermi levels in the p-n junction split, generating a photovoltage $V_{diode}$, which can be described by the classical diode equation.

This photovoltage is partitioned into two components within the device. One component raises the Fermi level (i.e., increases the electron energy) of the n-type semiconductor relative to that of the electrolyte, corresponding to the externally measured voltage V. The other component lowers the Fermi level of the p-type semiconductor and the protective layer relative to the electrolyte, corresponding to the overpotential $\eta$ that drives the interface electrochemical reaction. The interface charge transfer follows Butler-Volmer kinetics. For a single-electron transfer process, assuming symmetric charge transfer coefficients (0.5 for both forward and backward reactions), the following relation can be obtained[1]:

$$V(J) = V_{diode} - \eta = \frac{kT}{q} ln(\frac{J_L - J}{J_{sc}^0} + 1) - \frac{2RT}{F} arcsinh(\frac{J}{2J_{ex}}) \qquad (S1)$$

where k is the Boltzmann constant, T is the temperature, q is the elementary charge, $J_L$ is the photocurrent density, $J_{sc}^0$ is the reverse saturation current density of the semiconductor, R is the gas constant, F is the Faraday constant, and $J_{ex}$ is the exchange current density at the interface.

When the effects of series resistance $R_s$ and shunt resistance $R_{sh}$ are included, the current density J in the above expression is replaced by $J+(V+JR_s)/R_{sh}$, and the voltage V is replaced by $V+JR_s$.

Similarly, for a photocathode, the J-V relation becomes:

$$V(J) = -\frac{kT}{q} ln(\frac{J_L + J}{J_{sc}^0} + 1) + \frac{2RT}{F} arcsinh(\frac{J}{2J_{ex}}) \qquad (S2)$$

where the same substitutions can be applied to account for the effects of series and shunt resistances.

It should be noted that, in this work, the positive current direction is defined from the electrode toward the electrolyte. The applied voltage V is defined as the potential difference between the photoelectrode and the electrolyte, with upward band bending taken as the positive direction. This convention is consistent with common practices in both electrochemistry and solar cell research and is adopted throughout the following discussion.

## 1.2. Semiconductor-Electrolyte Junction Photoelectrodes

Existing studies on the current density-voltage (J-V) relationship of semiconductor-electrolyte junction (SEJ)

photoelectrodes often adopt relatively simplified formulations, where some key assumptions and intermediate steps are not explicitly presented[2,3]. This lack of transparency increases the difficulty in understanding the underlying physical logic. To improve the clarity and traceability of the derivation, a complete derivation of the governing equations is presented here.

Taking an n-type semiconductor as an example, as illustrated in Fig. S1(b), before contact, the Fermi level of the semiconductor, $E_{F,fb}^{sc}$ (where fb denotes the flat-band condition), is assumed to be higher than the Fermi level of the electrolyte, $E^{el}$, corresponding to the electrochemical potential of the electrolyte. Due to this difference in Fermi levels, electron transfer occurs upon contact, leading to a downward shift of the semiconductor Fermi level until equilibrium with $E^{el}$ is reached. During this process, band bending is formed near the semiconductor surface. Taking $x = x_D$ as the boundary, the region with band bending is defined as the space charge region (SCR), while the region with nearly flat bands in the bulk is referred to as the quasi-neutral region (QNR). The depletion approximation is adopted in this work, assuming zero net charge density in the QNR and an abrupt change in charge density at the SCR-QNR boundary, which is a classical assumption in solar cell theory.

Under illumination, the electron and hole quasi-Fermi levels in the semiconductor split. Due to the electric field in the SCR, photogenerated electrons tend to move toward the bulk, while photogenerated holes migrate toward the surface and accumulate near the interface. This increases the surface hole concentration and lowers the hole quasi-Fermi level, thereby providing the driving force for interface redox reactions.

Based on this framework, the J-V relationship for an n-type SEJ photoelectrode is derived. The derivation consists of three main steps[2,3]:

Establishing the expressions for interface electron and hole currents based on the Marcus-Gerischer theory;

Solving the hole continuity equation in the quasi-neutral region (QNR) to obtain the spatial distribution of holes, and determining the hole current at $x_D$;

Further solving the hole continuity equation in the space charge region (SCR) to derive the interface hole current and the relationship between the carrier concentrations at both ends of the SCR.

Finally, by combining the above results, the J-V relationship of the n-type SEJ photoelectrode can be obtained. The detailed derivation is provided as follows:

**Step 1: Interface electron and hole currents**

Assume that a single-electron redox reaction occurs at the interface:

$$Ox + e^- \rightleftarrows Red \tag{S3}$$

Assume that the interface carrier transfer follows a simple bimolecular reaction, the interface oxidation current density $J^+$ and reduction current density $J^-$ are respectively given by:

$$J^+ = qk_0 \int_{-\infty}^{+\infty} (1 - f(E)) \cdot D(E) \cdot D_{red}(E) \cdot dE \tag{S4}$$

$$J^- = -qk_0 \int_{-\infty}^{+\infty} f(E) \cdot D(E) \cdot D_{ox}(E) \cdot dE \tag{S5}$$

where $k_0$ is the rate constant, $f(E)$ is the Fermi–Dirac distribution function, and $D(E)$ is the density of states distribution in the semiconductor electrode, $D_{ox}(E)$ and $D_{red}(E)$ are the density of empty states and occupied states of the redox species in the electrolyte.

Electron transfer mainly occurs within an energy range of 1 kT near the conduction band or valence band, and electrons transfer through the conduction band while holes transfer through the valence band. Thus, for the current density for electron transfer from occupied states in the conduction band of the semiconductor to empty states in the electrolyte $J_e^-$, $E=E_C^s$ can be substituted into Eq. S5. In this case, $f(E_C^s)\cdot D(E_C^s)\cdot 1kT$ represents the surface electron concentration $n_s$. Therefore, the expression becomes:

$$J_e^- = -qk_0 n_s D_{ox}(E_C^s) \tag{S6}$$

Since $n_s$ varies with the applied potential, $J_e^-$ is a function of the voltage.

Analogously, for the current density for electron transfer from occupied states in the electrolyte to empty states in the conduction band of the semiconductor $J_e^+$, $E=E_C^s$ can be substituted into Eq. S4. In this case, since $f(E_C^s)$ is very small, $(1 - f(E_C^s))\cdot D(E_C^s)\cdot 1kT$ can be approximated as effective density of states in the conduction band $N_C$. Therefore, the expression becomes:

$$J_e^+ = qk_0 N_C D_{red}(E_C^S) \tag{S7}$$

Here, $J_e^+$ does not vary with the applied potential as $N_C$ is a constant.

Using the same idea, the current density for hole transfer from empty states in the electrolyte to occupied states in the valence band of the semiconductor is:

$$J_h^- = -qk_0 N_V D_{ox}(E_V^S) \tag{S8}$$

Here, $N_V$ is the effective density of states in the valence band. $J_h^-$ does not vary with the applied potential.

The current density for hole transfer from empty states in the valence band of the semiconductor to occupied states in the electrolyte is:

$$J_h^+ = qk_0 p_s D_{red}(E_V^S) \tag{S9}$$

Here, $p_s$ is the surface hole concentration. Since $p_s$ varies with the applied potential, $J_h^+$ is a function of the voltage.

At equilibrium, both the electron current $J_e$ and the hole current $J_h$ are zero. Therefore:

$$J_e(x=0) = J_{e,0}^- + J_{e,0}^+ = 0 \tag{S10}$$

$$J_h(x=0) = J_{h,0}^- + J_{h,0}^+ = 0 \tag{S11}$$

Accordingly, the exchange current densities for electrons and holes, $J_{ex,e}$ and $J_{ex,h}$, can be defined as:

$$J_{ex,e} = -J_{e,0}^- = J_{e,0}^+ = qk_0 N_C D_{red}(E_C^S) \tag{S12}$$

$$J_{ex,h} = -J_{h,0}^- = J_{h,0}^+ = qk_0 N_V D_{ox}(E_V^S) \tag{S13}$$

Under non-equilibrium conditions, the currents can be expressed as:

$$J_e(x=0) = J_{ex,e} \cdot \left(1 - \frac{n_s}{n_{s0}}\right) \tag{S14}$$

$$J_h(x=0) = J_{ex,h} \cdot \left(\frac{p_s}{p_{s0}} - 1\right) \tag{S15}$$

Here, $n_{s0}$ and $p_{s0}$ denote the equilibrium surface electron and hole concentrations, respectively.

**Step 2: Quasi-neutral region**

For an n-type semiconductor, hole transport is considered. The continuity equation is given by:

$$\frac{1}{q} \cdot \frac{dJ_h}{dx} = R - G \tag{S16}$$

where $J_h$ is the hole current density, and R and G are the recombination and generation rates, respectively. The hole current can be expressed as:

$$J_h = q\mu_h p\xi + qD_h \cdot \frac{dp}{dx} \tag{S17}$$

where $\mu_h$ is the hole mobility, p is the hole concentration, $\xi$ is the electric field, and $D_h$ is the hole diffusion coefficient. In the QNR, the electric field is negligible, and therefore the drift current vanishes; the current is dominated by diffusion.

The recombination rate R is written as:

$$R = \frac{p - p_0}{\tau_h} \tag{S18}$$

where $p_0$ is the equilibrium hole concentration in the semiconductor, and $\tau_h$ is the hole lifetime.

For monochromatic illumination with photon flux $\Phi_0(\lambda)$, under front-side illumination (i.e., light incident from the electrolyte into the semiconductor), the generation rate $G_F$ follows the Lambert-Beer law:

$$G_F(\lambda, x) = \alpha(\lambda) \cdot \big(1 - r(\lambda)\big) \cdot \Phi_0(\lambda) \cdot exp(-\alpha(\lambda)x) \tag{S19}$$

where $\alpha(\lambda)$ is the absorption coefficient and $r(\lambda)$ is the reflectance. Accordingly, the continuity equation becomes:

$$D_h \cdot \frac{d^2 p}{dx^2} - \frac{p - p_0}{\tau_h} + \alpha(1-r)\Phi_0 e^{-\alpha x} = 0 \tag{S20}$$

For a semi-infinite semiconductor, the boundary conditions are:

$$p(+\infty) = p_0 \tag{S21}$$

$$p(x_D) = p_D \tag{S22}$$

where $p_D$ is an unknown value to be determined later.

The hole distribution in the QNR is obtained as:

$$p(x) = p_0 + \left(p_D - p_0 - \frac{(1-r)\alpha\tau_h\Phi_0}{1-\alpha^2 {L_h}^2} \cdot e^{-\alpha x_D}\right) \cdot e^{-\frac{x-x_D}{L_h}} + \frac{(1-r)\alpha\tau_h\Phi_0}{1-\alpha^2 {L_h}^2} \cdot e^{-\alpha x} \quad (S23)$$

where $L_h$ is the hole diffusion length, defined by ${L_h}^2 = D_h\tau_h$.

The hole current at $x_D$ is then given by:

$$J_h(x_D) = qD_h \cdot \frac{dp}{dx}|_{x=x_D} = -J_h^0 \cdot \left(\frac{p_D}{p_0} - 1\right) + \frac{q\Phi_0(1-r)\alpha L_h \cdot e^{-\alpha x_D}}{1+\alpha L_h} \quad (S24)$$

where $J_h^0$ is the reverse saturation current of holes:

$$J_h^0 = \frac{qp_0 D_h}{L_h} \quad (S25)$$

**Step 3: Space charge region**

Recombination in the SCR is neglected. The hole continuity equation is integrated over the SCR:

$$\int_0^{x_D} \frac{1}{q} \cdot \frac{dJ_h}{dx} \cdot dx = \int_0^{x_D} -G(\lambda, x) \cdot dx \quad (S26)$$

Under front-side illumination, $G=G_F$. Therefore, the hole current at the interface (x = 0) is obtained as:

$$J_h(0) = J_h(x_D) + q\Phi_0(1-r)(1-e^{-\alpha x_D}) \quad (S27)$$

Next, the relationship between the carrier concentrations at the two ends of the SCR is considered. The hole current in the SCR consists of two components, diffusion and drift. Under moderate voltage or moderate illumination intensity, both components are large, and, to a leading order, the net current, i.e. the difference between the two, is negligible[4]. Physically, this means that holes in the SCR are nearly in local electrostatic equilibrium: the electric field tends to sweep holes in one direction by drift, while the strong spatial gradient in hole concentration drives diffusion in the opposite direction and these two fluxes almost cancel. Therefore, the following approximation typically holds:

$$-q\mu_h p\xi = qD_h \cdot \frac{dp}{dx} \quad (S28)$$

Using the Einstein relation to substitute $\mu_h$, and integrating within the SCR, considering that the forward direction of the current should be consistent with the electric field, it can be obtained:

$$p(0) = p_s = p(x_D) \cdot e^{\frac{q(V_D-V)}{kT}} \quad (S29)$$

Here, $V_D$ is the degree of band bending under equilibrium in dark, that is, the potential difference between the surface and the conduction band in the bulk. Surface defects are not considered here, and it is assumed that no Fermi level pinning occurs. Therefore, the positions of the surface conduction band $E_C^S$ and the surface valence band $E_V^S$ are fixed. It can be obtained:

$$V_D = E_{F,fb}^{sc} - E^{el} \quad (S30)$$

where $E_{F,fb}^{sc}$ is the Fermi level of the semiconductor and fb denotes the flat-band condition.

Therefore, the surface hole concentration under dark equilibrium $p_{s0}$ is:

$$p_{s0} = p_0 \cdot e^{\frac{qV_D}{kT}} \quad (S31)$$

Under illumination or applied bias, the surface hole concentration $p_s$ is shown in Eq. S29.

Similarly, the surface electron concentrations under equilibrium and non-equilibrium can be obtained, respectively:

$$n_{s0} = n_0 \cdot e^{-\frac{qV_D}{kT}} \quad (S32)$$

$$n_s = n(x_D) \cdot e^{\frac{q(V-V_D)}{kT}} \quad (S33)$$

where $n_0$ is the equilibrium electron concentration in the semiconductor, $n(x_D)$ is the electron concentration at $x=x_D$.

From $V_D$ and the depletion approximation, $x_D$ can be calculated as:

$$x_D = \sqrt{\frac{2\varepsilon_{sc}(V_D - V)}{qn_0}} \quad (S34)$$

where $\varepsilon_{sc}$ is the permittivity of the semiconductor.

Finally, by combining Eqs. S15, S22, S24, S27, S29 and S31, the hole current at the interface (x = 0) is obtained:

$$J_h(0) = \frac{J_h^0 \cdot \left(1 - e^{\frac{qV}{kT}}\right) + q\Phi_0(1-r)\left(1 - \frac{e^{-\alpha x_D}}{1+\alpha L_h}\right)}{1 + \frac{J_h^0}{J_{ex,h}} \cdot e^{\frac{qV}{kT}}} \tag{S35}$$

This is the hole current under monochromatic illumination. For the full-spectrum hole current, it is only necessary to integrate the second term of the expression. Define:

$$J_L = \int_0^{\lambda_g} q\Phi_0(1-r)\left(1 - \frac{e^{-\alpha x_D}}{1+\alpha L_h}\right) \cdot d\lambda \tag{S36}$$

where $\lambda_g$ is the cutoff wavelength, which can be calculated from the semiconductor bandgap $E_g$, Planck constant h, and the speed of light in vacuum c:

$$\lambda_g = \frac{hc}{E_g} \tag{S37}$$

Therefore, the full-spectrum hole current at the interface is:

$$J_h(0) = \frac{J_h^0 \cdot \left(1 - e^{\frac{qV}{kT}}\right) + J_L}{1 + \frac{J_h^0}{J_{ex,h}} \cdot e^{\frac{qV}{kT}}} \tag{S38}$$

**Final establishment of the J-V equation**

By combining Eqs. S14, S32 and S33, the electron current at the interface (x = 0) can be obtained. The small injection approximation is used here, that is, for an n-type semiconductor, electron injection under non-equilibrium does not affect the electron concentration. Therefore, $n(x_D)=n_0$. Thus, the electron current at the interface is:

$$J_e(0) = -J_{ex,e} \cdot \left(e^{\frac{qV}{kT}} - 1\right) \tag{S39}$$

The total current is the sum of the hole current and the electron current. Therefore, the current-voltage relationship of the n-type semiconductor SEJ can be obtained:

$$J(V) = \frac{J_h^0 \cdot \left(1 - e^{\frac{qV}{kT}}\right) + J_L}{1 + \frac{J_h^0}{J_{ex,h}} \cdot e^{\frac{qV}{kT}}} - J_{ex,e} \cdot \left(e^{\frac{qV}{kT}} - 1\right) \tag{S40}$$

If the effects of series resistance and shunt resistance are considered, it is only necessary to replace J in the above equation with $J+(V+JR_s)/R_{sh}$, and replace V with $V+JR_s$.

For a p-type semiconductor, the same method can be used to calculate the electron concentration in the QNR leading to:

$$J(V) = J_{ex,h} \cdot \left(e^{-\frac{qV}{kT}} - 1\right) - \frac{J_e^0 \cdot \left(1 - e^{-\frac{qV}{kT}}\right) + J_L}{1 + \frac{J_e^0}{J_{ex,e}} \cdot e^{-\frac{qV}{kT}}} \tag{S41}$$

If the effect of resistances is considered, the same substitution is applied.

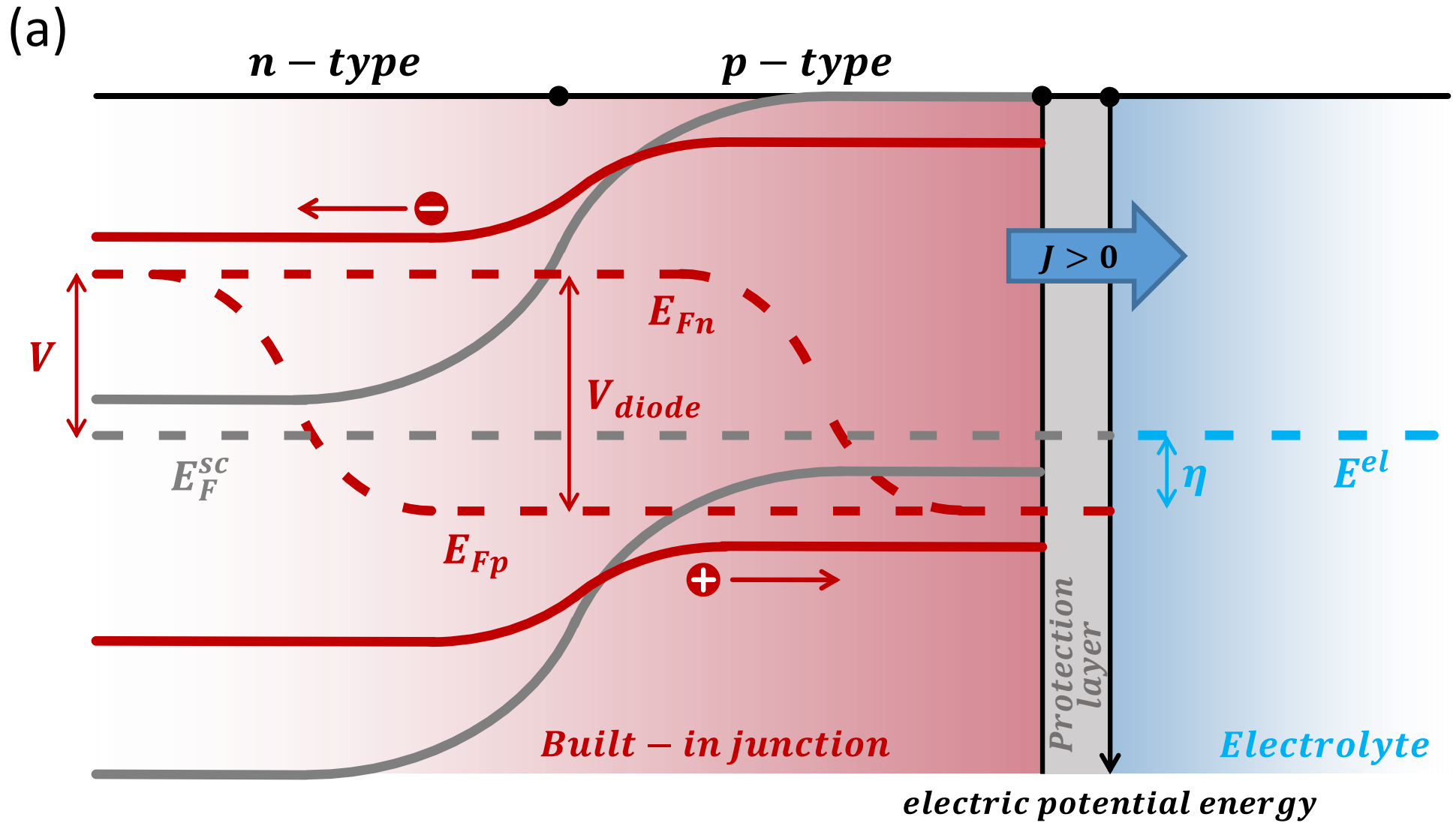

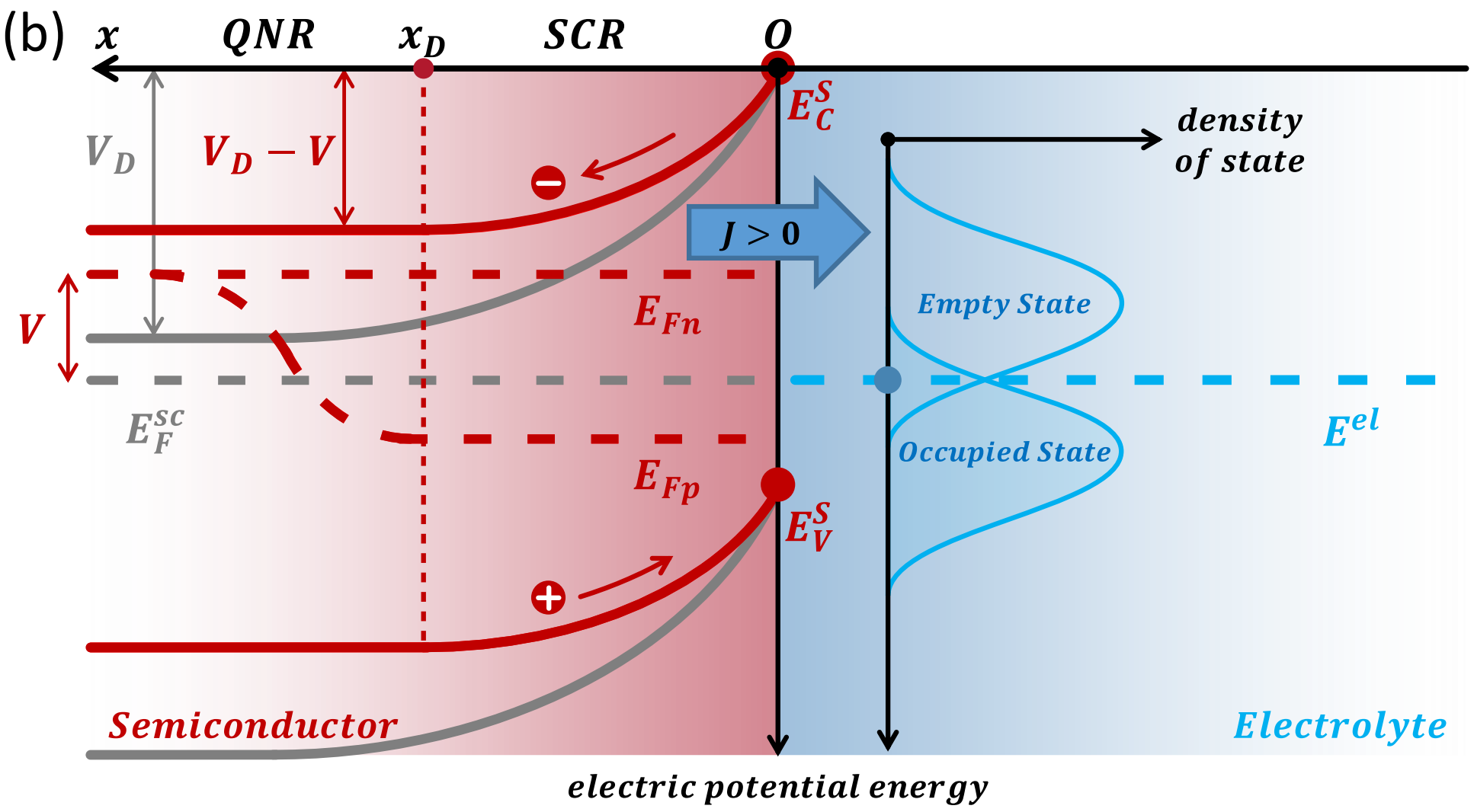

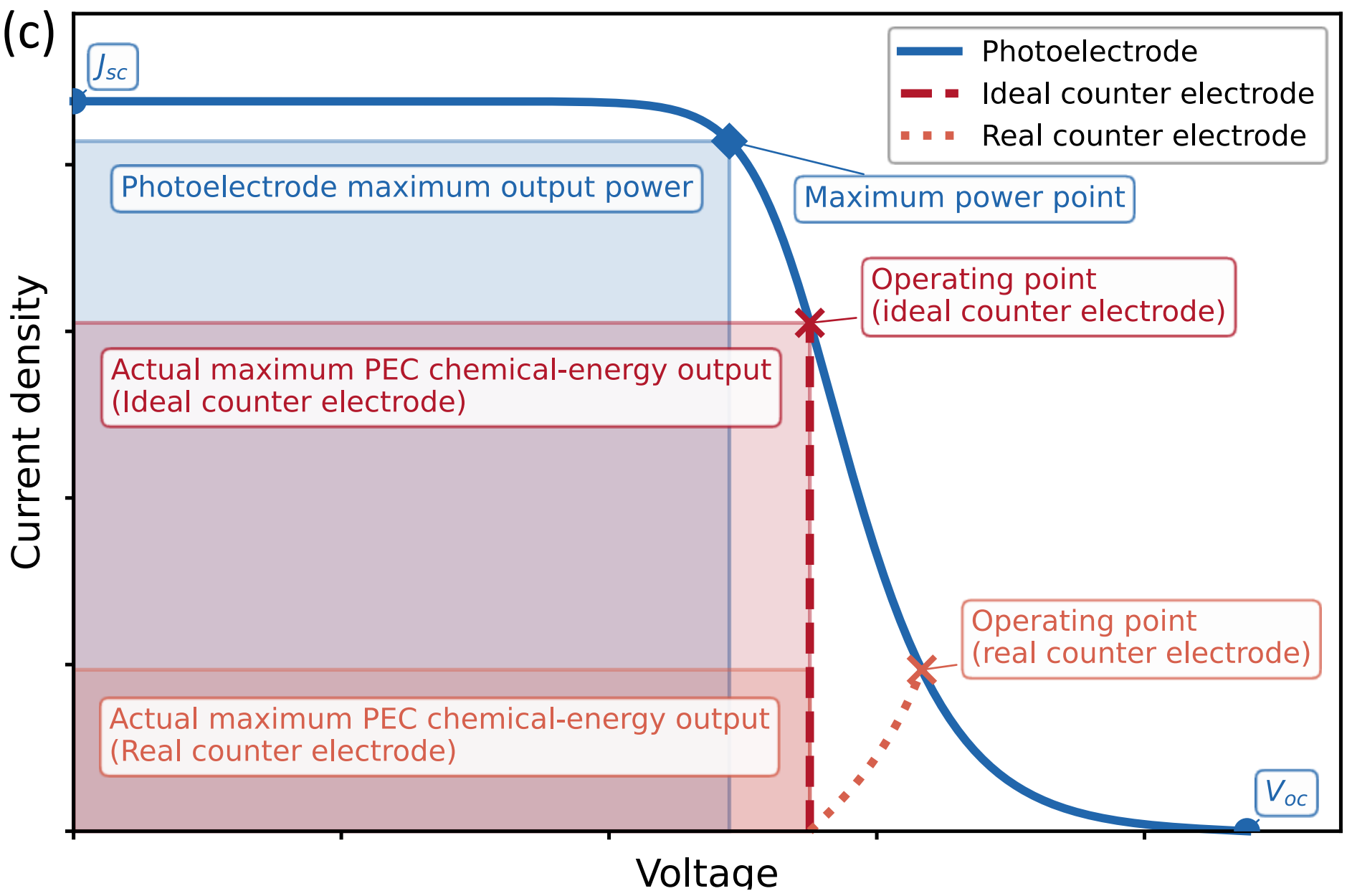

**Figure S1. Schematic illustration of photoelectrodes and their current density-voltage characteristics.**

(a) Schematic of a built-in junction photoelectrode;

(b) Schematic of a semiconductor-electrolyte junction photoelectrode;

(c) J-V characteristics of photoelectrode and counter electrode with operating points in PEC systems.

# 2. Semiconductor–electrolyte junction photoelectrode matching: detailed calculation of the interface exchange current densities of electrons and holes

The approximate calculations of Eqs. S4 and S5 in Section 1 are only generally valid within the semiconductor bandgap range. To perform a more accurate calculation, the densities of empty and occupied states of the redox species in the electrolyte can be expressed as follows:

$$D_{ox}(E) = \frac{c_{ox}}{\sqrt{4\pi\lambda_r kT}} \cdot \exp\left(-\frac{\left(E - E^{el} - \lambda_r\right)^2}{4\lambda_r kT}\right) \quad (S42)$$

$$D_{red}(E) = \frac{c_{red}}{\sqrt{4\pi\lambda_r kT}} \cdot \exp\left(-\frac{\left(E - E^{el} + \lambda_r\right)^2}{4\lambda_r kT}\right) \quad (S43)$$

where $c_{ox}$ and $c_{red}$ are the concentrations of the oxidized and reduced species, respectively, $\lambda_r$ is the reorganization energy, and $E^{el}$ is the electrolyte Fermi level.

Meanwhile, the density-of-states distribution in the semiconductor electrode can be divided into the conduction-band part $D_C$ and the valence-band part $D_V$:

$$D_C(E) = (m_e^*)^{\frac{3}{2}} \cdot \sqrt{E - E_C^S} \quad (S44)$$

$$D_V(E) = (m_h^*)^{\frac{3}{2}} \cdot \sqrt{E_V^S - E} \quad (S45)$$

where $m_e^*$ and $m_h^*$ are the effective masses of electrons and holes, respectively.

Accordingly, $J_e^+$ and $J_h^-$ can be further expressed as:

$$J_e^+ = qk_0(m_e^*)^{\frac{3}{2}} \cdot \frac{c_{red}}{\sqrt{4\pi\lambda_r kT}} \cdot \int_{E_C^S}^{+\infty} \left(1 - f(E)\right) \cdot \sqrt{E - E_C^S} \cdot \exp\left(-\frac{\left(E - E^{el} + \lambda_r\right)^2}{4\lambda_r kT}\right) \cdot dE \quad (S46)$$

$$J_h^- = -qk_0(m_h^*)^{\frac{3}{2}} \cdot \frac{c_{ox}}{\sqrt{4\pi\lambda_r kT}} \cdot \int_{-\infty}^{E_V^S} f(E) \cdot \sqrt{E_V^S - E} \cdot \exp\left(-\frac{\left(E - E^{el} - \lambda_r\right)^2}{4\lambda_r kT}\right) \cdot dE \quad (S47)$$

According to the definitions in Section 1, $J_{ex,e}$ corresponds to $J_e^+$, whereas $J_{ex,h}$ corresponds to $-J_h^-$. Therefore, the reduced $J_{ex,e}$ and $J_{ex,h}$ can be defined as:

$$reduced\, J_{ex,e} = \frac{J_e^+}{qk_0(m_e^*)^{\frac{3}{2}} c_{red}} = \frac{1}{\sqrt{4\pi\lambda_r kT}} \cdot \int_{E_C^S}^{+\infty} \left(1 - f(E)\right) \cdot \sqrt{E - E_C^S} \cdot \exp\left(-\frac{\left(E - E^{el} + \lambda_r\right)^2}{4\lambda_r kT}\right) \cdot dE \quad (S48)$$

$$reduced\, J_{ex,h} = \frac{-J_h^-}{qk_0(m_h^*)^{\frac{3}{2}} c_{ox}} = \frac{1}{\sqrt{4\pi\lambda_r kT}} \cdot \int_{-\infty}^{E_V^S} f(E) \cdot \sqrt{E_V^S - E} \cdot \exp\left(-\frac{\left(E - E^{el} - \lambda_r\right)^2}{4\lambda_r kT}\right) \cdot dE \quad (S49)$$

As shown in Fig. S2(a), within the semiconductor bandgap range, $J_{ex,e}$ and $J_{ex,h}$ are nearly independent of the semiconductor Fermi level, which further validates the rationality of the approximations used in Section 1.

Fig. S2(b,c) show the variations of $J_{ex,e}$ and $J_{ex,h}$ with the electrolyte Fermi level and the reorganization energy in p-type semiconductors. The observed trends are consistent with the conclusions obtained for n-type semiconductors in the main text.

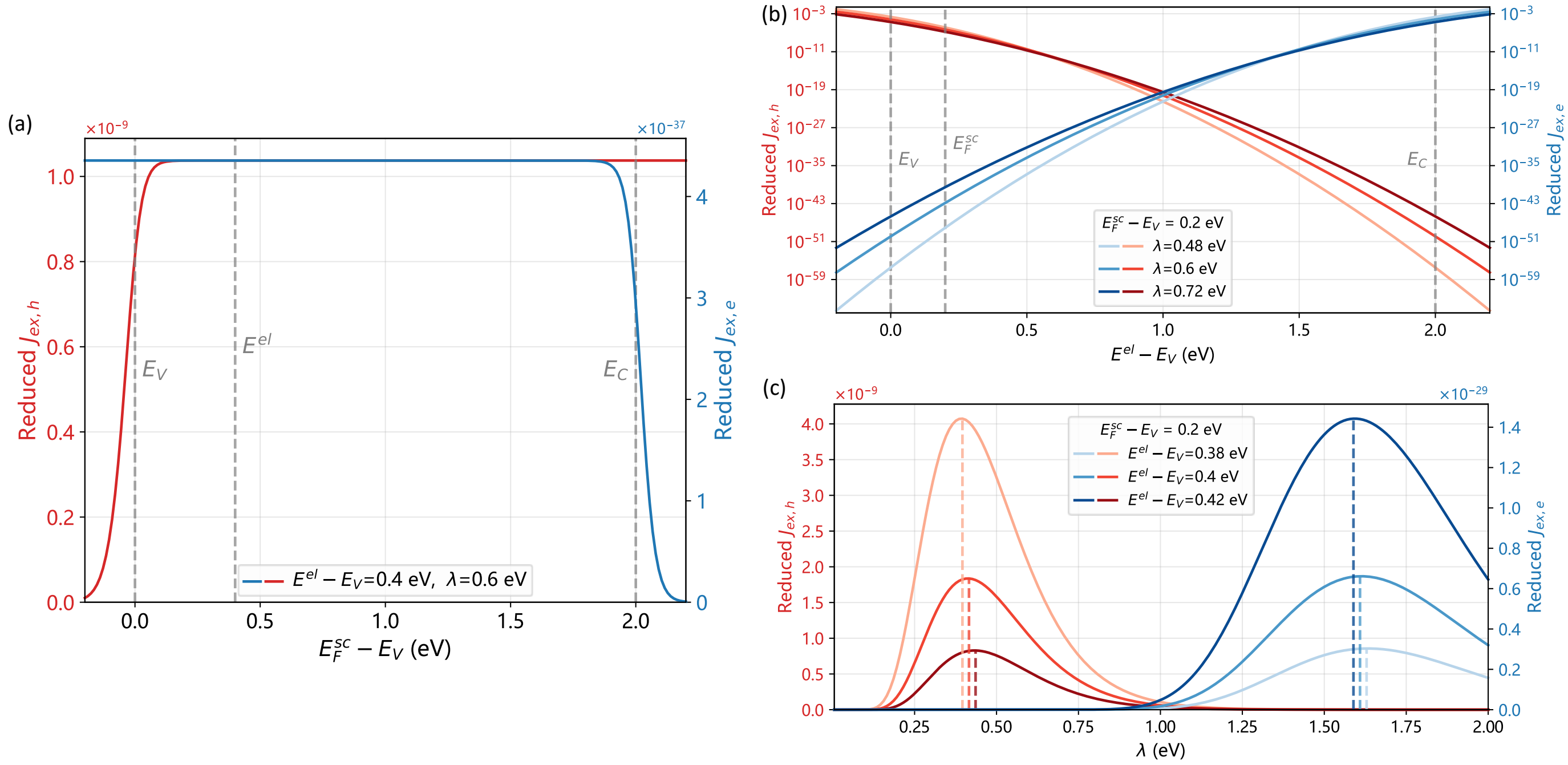

**Figure S2. Dependence of the reduced interface exchange current densities of electrons and holes on the semiconductor Fermi level, electrolyte Fermi level, and reorganization energy.**

# 3. Performance Metrics of photoelectrodes and PEC systems

## 3.1. Performance of Photoelectrodes

Once the J-V relationship of a photoelectrode is obtained, its device performance can be evaluated accordingly. Following the analysis commonly used for solar cells, the main performance metrics of a photoelectrode can be directly determined from the J-V curve (Fig. S1(c)), including the following parameters:

Short-circuit current density ($J_{SC}$): the current density when the potential difference between the photoelectrode and the electrolyte is zero;

Open-circuit voltage ($V_{OC}$): the potential difference between the photoelectrode and the electrolyte when the current density is zero;

Maximum power point (MPP): the point on the J-V curve where the power density $P=J\cdot V$ reaches its maximum value. The corresponding current density and voltage are denoted as $J_{MPP}$ and $V_{MPP}$, respectively;

Maximum energy output efficiency of the photoelectrode (photoelectrode efficiency), defined as:

$$\eta_{PE} = \frac{J_{MPP} \cdot V_{MPP}}{P_{in}} \tag{S50}$$

where $P_{in}$ is the incident solar power density. The numerator represents the output power density of the photoelectrode at the maximum power point, corresponding to the blue shaded area shown in Fig. S1(c);

Fill factor (FF), defined as:

$$FF = \frac{J_{MPP} \cdot V_{MPP}}{J_{SC} \cdot V_{OC}} \tag{S51}$$

which represents the ratio between the maximum output power and the ideal rectangular power $J_{SC}\cdot V_{OC}$, and characterizes the deviation of the device performance from the ideal case.

## 3.2. Performance of Photoelectrochemical Cells

A PEC cell consists of a photoelectrode and a counter electrode. Taking the reaction potential at the photoelectrode side as the reference, the reaction potential of the counter electrode corresponds to the reaction potential difference ΔU. The J-V relationship of the counter electrode follows the Butler-Volmer equation. For a single-electron transfer process, assuming equal charge-transfer coefficients of 0.5 for the forward and backward reactions, the current density can be written as:

$$J = J_{ex,c} \cdot (e^{\frac{q(V-\Delta U)}{2kT}} - e^{-\frac{q(V-\Delta U)}{2kT}}) \tag{S52}$$

where $J_{ex,c}$ is the interface exchange current density of the counter electrode.

The operating point of a PEC cell is determined by the intersection of the J-V curves of the photoelectrode and the counter electrode. Taking a photoanode as an example (Fig. S1(c)), in the ideal case where $J_{ex,c}$ is sufficiently large, the polarization of the counter electrode can be neglected and its J-V characteristic approaches a vertical line. In practical cases, however, a finite $J_{ex,c}$ leads to counter electrode polarization, resulting in a curved J-V response and a shift of the operating point. The corresponding operating current density is denoted as $J_{OP}$. The output power of the PEC cell can therefore be expressed as

$$P_{PEC,out} = J_{OP} \cdot \Delta U \tag{S53}$$

The energy conversion efficiency of the PEC cell (PEC efficiency) is then defined as

$$\eta_{PEC} = \frac{P_{PEC,out}}{P_{in}} \tag{S54}$$

Furthermore, a utilization factor of the photoelectrode, γ, can be introduced to quantify how effectively the intrinsic output capability of the photoelectrode is utilized under actual PEC cell operating conditions:

$$\gamma = \frac{\eta_{PEC}}{\eta_{PE}} \tag{S55}$$

In summary, based on the J-V relationships established above, both the performance of the photoelectrode and the overall performance of the PEC cell can be quantitatively calculated and predicted.

# 4. Quantitative Influence of Parameters on Efficiency

A quantitative analysis based on the established model is conducted to evaluate their effects on photoelectrode and PEC efficiencies and to extract general optimization principles. Representative semiconductor systems with bandgaps of 1.12 eV and 2.1 eV are considered for BIJ and SEJ PEC systems, respectively.

For BIJ PEC systems, the effects of semiconductor parameters and interface parameters on photoelectrode efficiency are first examined, as shown in Fig. S3(a) and (b), respectively. The bandgap exhibits two local maxima in efficiency at ~1.1 eV and ~1.3 eV, with a theoretical limit exceeding 33%. This behavior arises from the trade-off between incomplete absorption at large bandgaps and increased thermalization losses at small bandgaps, consistent with solid-state solar cells. The bandgap therefore fundamentally determines the upper efficiency limit and serves as a primary criterion for material selection.

As $\beta$ and $\theta$ approach ideal values, the efficiency increases. The effect of $\beta$ is moderate, with a one-order-of-magnitude decrease resulting in only ~2% efficiency loss. In contrast, $\theta$ has a much stronger impact: when reduced from 1 to 0.1, the efficiency drops below 5%. This indicates that improving effective photocarrier generation and collection is more critical than extending carrier lifetime.

A lower series resistance $R_s$ reduces voltage loss, while a higher shunt resistance $R_{sh}$ suppresses parasitic currents and improves efficiency. Increasing $J_{ex}$ enhances charge transfer kinetics. However, the efficiency saturates beyond a threshold, indicating that interface reactions are no longer limiting and that the effective regime for interface optimization is restricted.

The effect of counter electrode parameters on utilization factor $\gamma$ of photoelectrode is shown in Fig. S3(c). The $\Delta U$ exhibits a clear optimum corresponding to the MPP voltage, indicating that the operating point is determined by J-V matching between the two electrodes. Increasing $J_{ex,c}$ improves interface kinetics at the counter electrode and shifts the operating point toward the ideal condition.

For SEJ PEC systems, the trends of most parameters are consistent with BIJ systems, except for $J_{ex,h}$ and $J_{ex,e}$ (Fig. S3(d–f)). Under photoanode operation, interface reactions are dominated by holes. A higher $J_{ex,h}$ therefore enhances interface charge transfer kinetics, facilitating hole injection into the electrolyte and improving photoelectrode efficiency. In contrast, a lower $J_{ex,e}$ suppresses reverse and parasitic reactions, thereby improving efficiency. Under photocathode condition, these roles are correspondingly reversed.

Notably, as discussed above, Eqs. S12 and S13 show that the magnitudes of $J_{ex,e}$ and $J_{ex,h}$ are jointly determined by the intrinsic properties of both the semiconductor and the electrolyte. Therefore, these two parameters not only characterize interface charge transfer kinetics, but also reflect the degree of matching between the semiconductor and the electrolyte, and can serve as important criteria for material selection and system design.

Overall, among all parameters, only the semiconductor bandgap $E_g$ and the reaction potential difference $\Delta U$ exhibit clear optimal values, corresponding to the intrinsic properties of the photoelectrode material and the electrolyte reaction system, respectively, and can therefore serve as core parameters for material selection and system design. The effects of the remaining parameters on efficiency are largely monotonic and can be continuously improved through targeted optimization strategies.

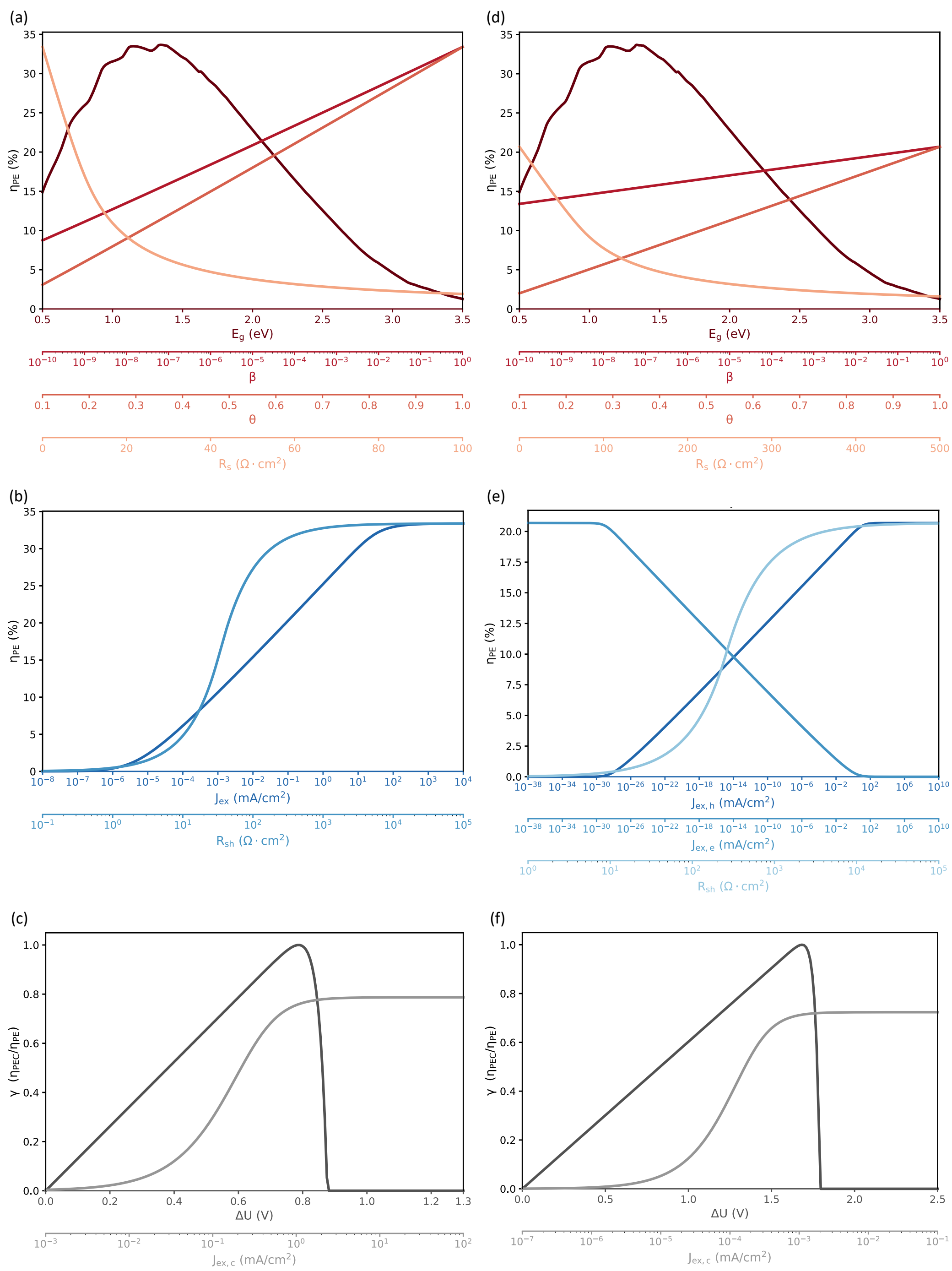

**Figure S3. Quantitative influence of parameters on efficiency.**

(a) Influence of semiconductor parameters on built-in junction photoelectrode efficiency.

(b) Influence of interface parameters on built-in junction photoelectrode efficiency.

(c) Influence of counter electrode parameters on built-in junction photoelectrode utilization factor in a photoelectrochemical cell.

(The semiconductor bandgap is fixed at 1.12 eV. Except for the parameter under investigation, all other parameters are set to ideal values defined in Fig. 1(d). For the light grey curve in Fig. S3(c), ΔU=0.6 V.)

(d) Influence of semiconductor parameters on semiconductor-electrolyte junction photoelectrode efficiency.

(e) Influence of interface parameters on semiconductor-electrolyte junction photoelectrode efficiency.

(f) Influence of counter electrode parameters on semiconductor-electrolyte junction photoelectrode utilization factor in a photoelectrochemical cell.

(The semiconductor bandgap is fixed at 2.1 eV. Except for the parameter under investigation, all other parameters are set to ideal values defined in Fig. 1(e). For the light grey curve in Fig. S3(f), ΔU=1.2 V.)

# 5. Shapley allocation method[5]

Let f be a function of n variables:

$$f: \mathbb{R}^n \to \mathbb{R} \tag{S56}$$

Let the initial and final states be:

$$\overrightarrow{x^{(1)}} = \left(x_1^{(1)}, x_2^{(1)}, \dots, x_n^{(1)}\right) \tag{S57}$$

$$\overrightarrow{x^{(2)}} = \left(x_1^{(2)}, x_2^{(2)}, \dots, x_n^{(2)}\right) \tag{S58}$$

Define the index set:

$$N = \{1,2, \dots, n\} \tag{S59}$$

For any subset $S \subseteq N$, define the mixed-state vector $\overrightarrow{x^{(1,2)}(S)} \in \mathbb{R}^n$ componentwise as:

$$\forall k \in N, (\overrightarrow{x^{(1,2)}(S)})_k = \begin{cases} x_k^{(2)}, k \in S \\ x_k^{(1)}, k \notin S \end{cases} \tag{S60}$$

Then, for each $i \in N$, the Shapley contribution of variable $x_i$ is defined as:

$$\phi_i = \sum_{S \subseteq N \setminus \{i\}} \frac{|S|!\,(n - |S| - 1)!}{n!} \left[ f\left(\overrightarrow{x^{(1,2)}(S \cup \{i\})}\right) - f\left(\overrightarrow{x^{(1,2)}(S)}\right) \right] \tag{S61}$$

In this study, the function f represents the efficiency of the photoelectrode, with its variables defined as the set of parameters introduced in Table 2 in the main text, excluding the bandgap. As these parameters change from one configuration to another, the efficiency of the photoelectrode correspondingly varies. The Shapley attribution method can be employed to quantitatively evaluate the contribution of each parameter to this change.

# 6. Selection of State-of-the-Art Studies, Curve Fitting, and Parameter Extraction for Conventional and Emerging Photoelectrode Materials

This study selects representative works on single-junction photoelectrodes from nine materials reported up to the end of 2025 (photovoltaic grade: Si, perovskite, CIGS, CZTS, $Sb_2Se_3$; non-photovoltaic grade: $\alpha$-$Fe_2O_3$, $Ta_3N_5$, $Cu_2O$, $BiVO_4$), and further screens those with the highest short circuit photocurrent density.

The J-V curves of the best-performing photoelectrodes in these works are fitted using the model developed in the main text, and the corresponding parameters are extracted (see Table S1 and S2). In addition, two representative studies on Si and $Ta_3N_5$ are analyzed in detail, where the variations in J-V curves for photoelectrodes with different structures are compared, and the fitting is performed accordingly to obtain the parameters (see Table S3 and S4).

First, starting from the J-V curves of the photoelectrodes reported in the selected literature, the corresponding data points are extracted using the online tool PlotDigitizer[6]. On this basis, the J-V curves of built-in junction photoelectrodes and semiconductor-electrolyte junction photoelectrodes are fitted using our self-developed Python program to obtain the relevant parameters. The goodness of fit is evaluated using the coefficient of determination $R^2$, where a value closer to 1 indicates better fitting performance. The definition of $R^2$ is given as follows:

$$R^2 = 1 - \frac{\sum(J_{data} - J_{model})^2}{\sum(J_{data} - \overline{J_{data}})^2} \qquad (S62)$$

Here, $J_{data}$ represents the experimentally measured current density data points extracted from the literature, $J_{model}$ denotes the current density calculated from the model and its parameters under the same voltage conditions, and $\overline{J_{data}}$ is the mean value of all experimental current density data points.

In the specific implementation of the fitting procedure, random sampling is first performed for a coarse global search to identify potential optimal regions, followed by local refinement around the best solution. This refinement process is constrained by both a maximum number of refinement iterations and a minimum threshold for the search step size; the computation is terminated when either the maximum number of iterations is reached or the step sizes in all parameter directions are reduced below the predefined lower bound. By applying a consistent fitting procedure and evaluation criterion to all samples, the comparability and objectivity of the results are ensured.

All fitted curves and the corresponding $R^2$ values are shown in Fig. S4 and S5. The results indicate that nearly all fittings yield $R^2$ values not lower than 0.995, demonstrating that the model provides an excellent description of the experimental data.

**Table S1: Fitted Parameters for Representative Built-in Junction Photoelectrodes**

| Photoelectrode materials | β | θ | $J_{ex}$ (mA/cm$^2$) | $R_s$ (Ω*cm$^2$) | $R_{sh}$ (Ω*cm$^2$) |
|---|---|---|---|---|---|
| Si[7] | $4.075\times10^{-6}$ | 0.9515 | $1.399\times10^{2}$ | 4.927 | $3.573\times10^{2}$ |
| Perovskite[8] | $6.863\times10^{-2}$ | 0.8119 | $3.756\times10^{-1}$ | 2.540 | $3.601\times10^{2}$ |
| CIGS[9] | $2.176\times10^{-2}$ | 0.8101 | $4.836\times10^{-1}$ | 3.731 | $1.834\times10^{2}$ |
| CZTS[10] | $1.000\times10^{-1}$ | 0.9361 | $4.167\times10^{-2}$ | 6.764 | $1.696\times10^{2}$ |
| $Sb_2Se_3$[11] | $1.982\times10^{-8}$ | 0.9247 | $4.961\times10^{-1}$ | 7.327 | $2.156\times10^{2}$ |
| α-$Fe_2O_3$[12] | $6.032\times10^{-12}$ | 0.9207 | $1.513\times10^{-7}$ | 12.23 | $2.671\times10^{3}$ |
| Best parameters | $1.000\times10^{-1}$ | 0.9515 | $1.399\times10^{2}$ | 2.540 | $2.671\times10^{3}$ |

**Table S2: Fitted Parameters for Representative Semiconductor-electrolyte Junction Photoelectrodes**

| Photoelectrode materials | β | θ | $J_{ex,h}$ (mA/cm$^2$) | $J_{ex,e}$ (mA/cm$^2$) | $R_s$ (Ω*cm$^2$) | $R_{sh}$ (Ω*cm$^2$) |
|---|---|---|---|---|---|---|
| $Ta_3N_5$[13] | $8.066\times10^{-6}$ | 1.000 | $4.039\times10^{-13}$ | $1.651\times10^{-22}$ | 35.60 | $4.443\times10^{2}$ |
| $Cu_2O$[14] | $1.199\times10^{-8}$ | 1.000 | $3.515\times10^{-11}$ | $7.310\times10^{-9}$ | 30.05 | $5.285\times10^{2}$ |
| $BiVO_4$[15] | $1.068\times10^{-8}$ | 0.8931 | $1.580\times10^{-12}$ | $1.005\times10^{-25}$ | 63.76 | $5.166\times10^{3}$ |
| Best parameters | $8.066\times10^{-6}$ | 1.000 | $3.515\times10^{-11}$ | $1.005\times10^{-25}$ | 30.05 | $5.166\times10^{3}$ |

**Table S3: Fitted Parameters of Si Photoelectrodes with Different Structures**

| Photoelectrode materials | β | θ | $J_{ex}$ (mA/cm$^2$) | $R_s$ (Ω*cm$^2$) | $R_{sh}$ (Ω*cm$^2$) |
|---|---|---|---|---|---|
| $n^+np^+$-Si +Ni[7] | $1.102\times10^{-6}$ | 0.9457 | $1.072\times10^{-1}$ | 7.037 | $7.037\times10^{2}$ |
| $n^+np^+$-Si +Ni+ $MoS_2/Ni_3S_2$[7] | $4.075\times10^{-6}$ | 0.9515 | $1.399\times10^{2}$ | 4.927 | $3.573\times10^{2}$ |

**Table S4: Fitted Parameters of $Ta_3N_5$ Photoelectrodes with Different Structures**

| Photoelectrode materials | β | θ | $J_{ex,h}$ (mA/cm$^2$) | $J_{ex,e}$ (mA/cm$^2$) | $R_s$ (Ω*cm$^2$) | $R_{sh}$ (Ω*cm$^2$) |
|---|---|---|---|---|---|---|
| $Ta_3N_5$[13] | $2.932\times10^{-11}$ | 0.8915 | $3.526\times10^{-16}$ | $7.956\times10^{-11}$ | 237.8 | $2.419\times10^{12}$ |
| $Ta_3N_5$-$AlO_x$/Fh-$NiFeO_x$/$CeO_x$[13] | $8.066\times10^{-6}$ | 1.000 | $4.039\times10^{-13}$ | $1.651\times10^{-22}$ | 35.60 | $4.443\times10^{2}$ |

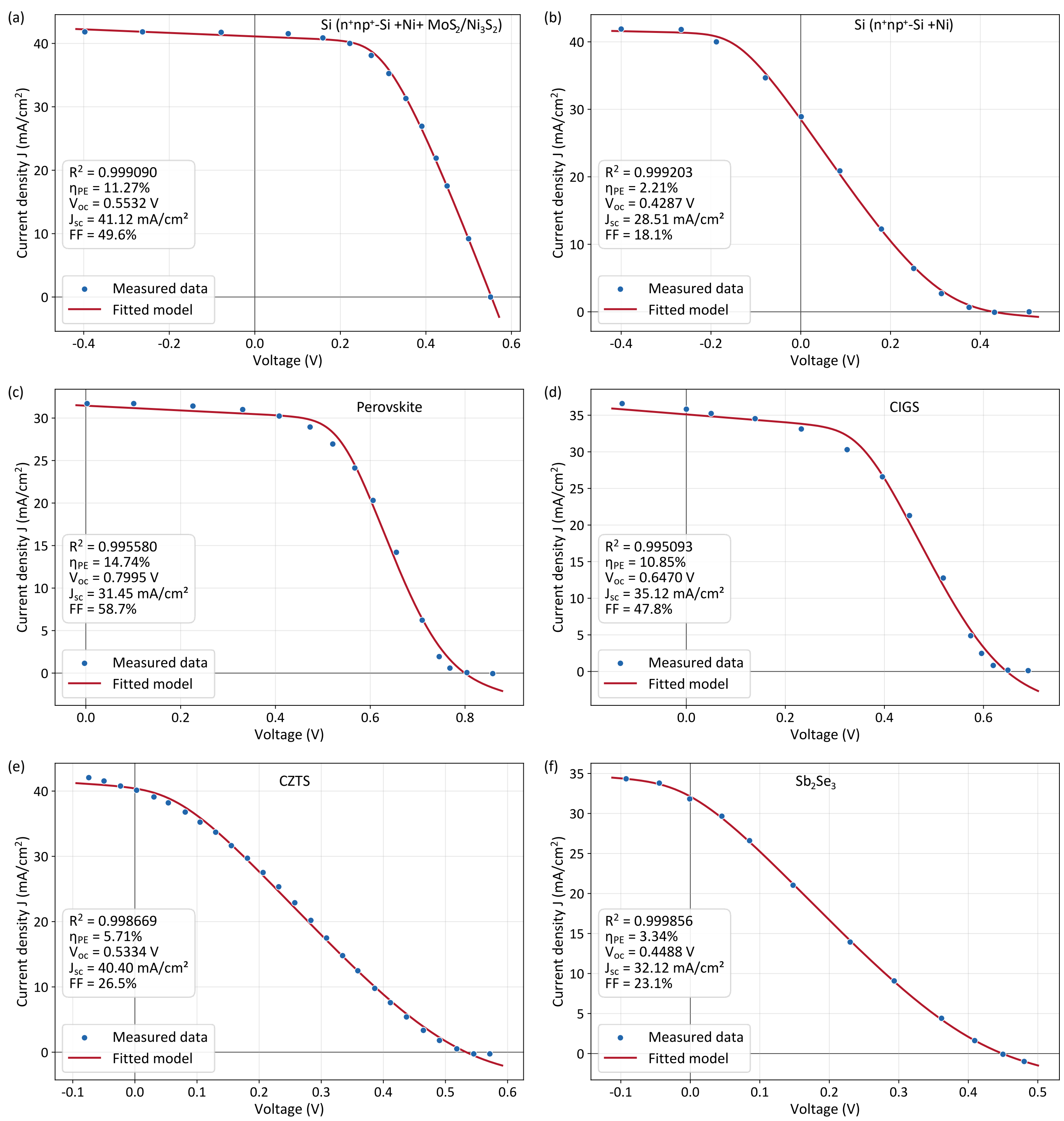

**Figure S4. J-V Curve Fitting of Representative Photoelectrodes Based on PV-Grade Materials.**

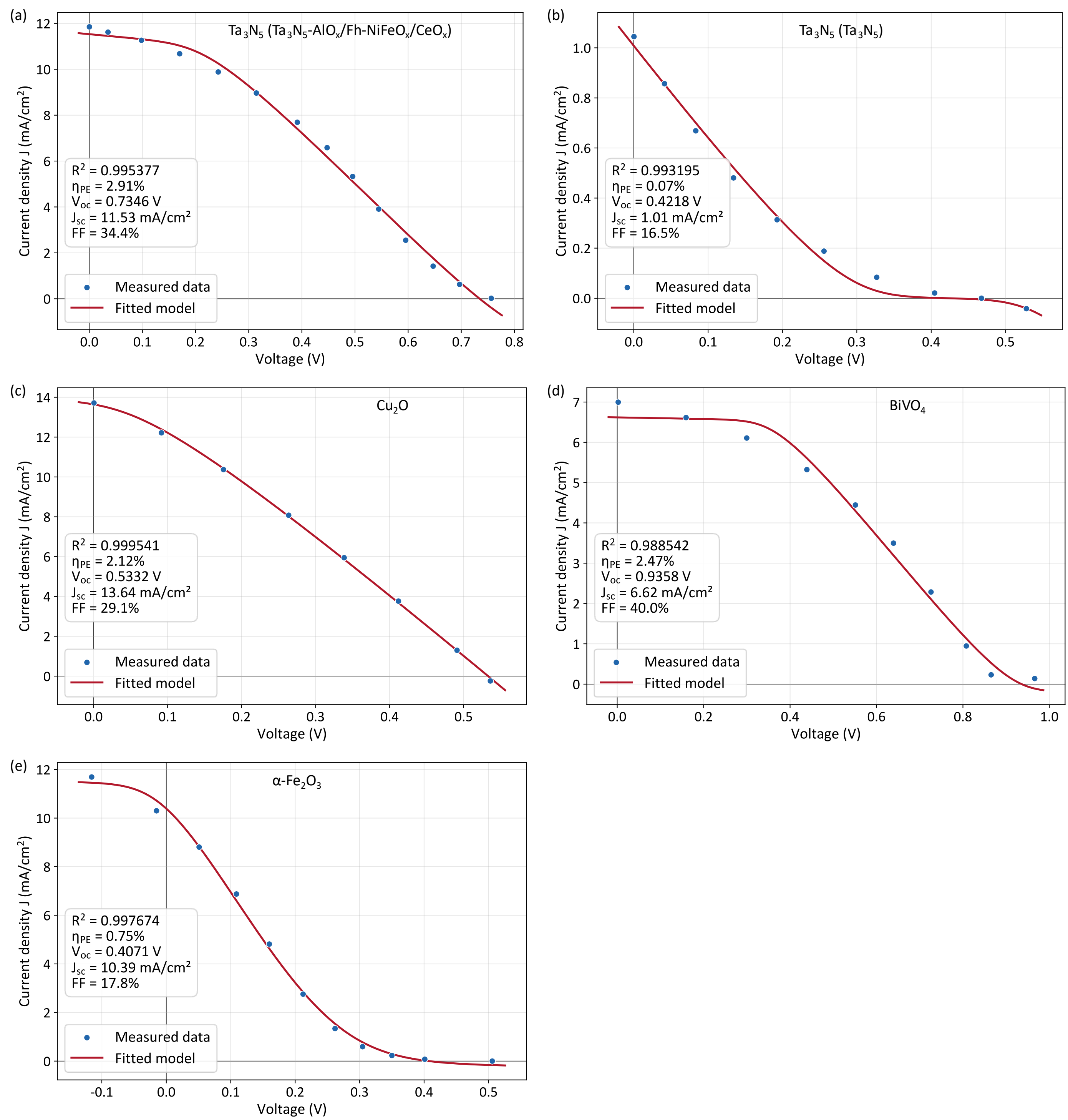

**Figure S5. J-V Curve Fitting of Representative Photoelectrodes Based on Non-PV Grade Materials.**

# 7. Applicability of the PEC Model in Solar $CO_2$ Reduction, Solar Redox Flow Batteries, and Solar Ammonia Synthesis

The rationality of the model has been validated in the main text and Section 6 of the Supplementary Information from two aspects:

1. through the $R^2$ values obtained from curve fitting;
2. through the consistency between the model-predicted energy loss analysis and the experimental conclusions.

However, the previously selected high-photocurrent studies are mainly focused on PEC systems for solar water splitting. To further verify the applicability of the model to other PEC applications, this work selects three representative studies in recent years on solar $CO_2$ reduction, solar redox flow batteries (SRFB), and solar ammonia synthesis. The J-V curves of the photoelectrodes in these studies are fitted, and the corresponding energy loss analysis is carried out.

The results are evaluated again from the two aspects mentioned above, further demonstrating the applicability and validity of the model in these three fields.

For the solar $CO_2$ reduction system, an $n^+$p-Si built-in junction is employed as the photoelectrode, and the J-V curves of three configurations are compared: bare $n^+$p-Si, $n^+$p-Si with Au catalyst, and $n^+$p-Si with a nanoporous Au mesh[16]. Based on the model presented in this work, the three curves are fitted, and the results are shown in Fig. S6(a-c), indicating high fitting accuracy.

The fitting range is further extended to the negative voltage region, as the $CO_2$ reduction reaction exhibits sluggish kinetics, and the positive voltage region cannot adequately represent the characteristics of the curves. The fitted parameters are listed in Table S5.

On this basis, an energy loss analysis is carried out, and the results are shown in Fig. S7(a). As can be seen from Table S5 and Fig. S7(a), the introduction of the catalyst improves the interface, which further leads to a significant increase in the carrier lifetime in the bulk. In addition, the introduction of the catalyst partially blocks the incident light, resulting in a reduction in $\theta$, which is also captured by the model. It should be noted that, due to the small current in the positive voltage region, the corresponding parameter variations are not very pronounced in the energy loss analysis but very clear in the table S5. These results are consistent with those reported in the literature.

In summary, the proposed model is applicable to the solar $CO_2$ reduction system.

For the SRFB system, an $n^+$p-Si built-in junction is employed as the photoelectrode absorber[17], with Ti conductive and $TiO_2$ protective layers subsequently deposited on top. The J-V characteristics with and without a carbon (C) catalyst are compared.

Both curves are fitted using the proposed model, and the fitting results, together with the corresponding $R^2$ values, are shown in Fig. S6(d,e), demonstrating high fitting accuracy. The extracted parameters are listed in Table S6.

Based on this, an energy loss analysis is further conducted, and the resulting donut charts are presented in Fig. S7(b). The results indicate that the introduction of the catalyst accelerates interface kinetics, thereby extending the carrier lifetime. Consequently, both interface and bulk losses are reduced. In addition, the decrease in light absorption induced by the catalyst layer is also captured by the model.

Overall, the proposed model is also applicable to the SRFB system.

For the solar ammonia synthesis, $Sb_2Se_3$ is used as the semiconductor material, forming a semiconductor-electrolyte junction with a potassium nitrate-containing electrolyte, and a $TiO_2$ layer is introduced as a protective layer[18]. At the same time, the J-V curves with and without Co&Cu catalysts are compared.

Based on the established model, the two curves are fitted, and the results are shown in Fig. S6(f,g), indicating a high fitting accuracy. The corresponding parameters are listed in Table S7.

On this basis, an energy loss analysis is further carried out, and the results are shown in Fig. S7(c). It can be seen that, after introducing the catalyst, the interface reaction kinetics are significantly enhanced, while the resistance to carrier transport is reduced, as reflected by the decrease in series resistance.

Thus, the proposed model is also applicable to the solar ammonia synthesis system.

**Table S5 Fitted Parameters for Si Photoelectrodes for Solar $CO_2$ Reduction**

| Photoelectrode materials | β | θ | $J_{ex}$ (mA/cm$^2$) | $R_s$ (Ω*cm$^2$) | $R_{sh}$ (Ω*cm$^2$) |
|---|---|---|---|---|---|
| n$^+$p-Si[16] | $7.913\times10^{-10}$ | 0.4438 | $4.360\times10^{-7}$ | 47.42 | $9.897\times10^{2}$ |
| n$^+$p-Si+Au[16] | $9.560\times10^{-7}$ | 0.2231 | $3.689\times10^{-4}$ | 59.62 | $6.533\times10^{3}$ |
| n$^+$p-Si+nanoporous Au mesh | $4.800\times10^{-7}$ | 0.2004 | $5.512\times10^{-3}$ | 67.21 | $6.246\times10^{2}$ |

**Table S6: Fitted Parameters for Si Photoelectrodes for Solar Redox Flow Batteries**

| Photoelectrode materials | β | θ | $J_{ex}$ (mA/cm$^2$) | $R_s$ (Ω*cm$^2$) | $R_{sh}$ (Ω*cm$^2$) |
|---|---|---|---|---|---|
| n$^+$p-Si+Ti+$TiO_2$[17] | $1.049\times10^{-8}$ | 0.6596 | $2.775\times10^{-2}$ | 10.63 | $2.079\times10^{4}$ |
| n$^+$p-Si+Ti+$TiO_2$+C[17] | $1.661\times10^{-4}$ | 0.2810 | $2.948\times10^{-1}$ | 3.574 | $1.189\times10^{3}$ |

**Table S7: Fitted Parameters for $Sb_2Se_3$ Photoelectrodes for Solar Ammonia Synthesis**

| Photoelectrode materials | β | θ | $J_{ex,h}$ (mA/cm$^2$) | $J_{ex,e}$ (mA/cm$^2$) | $R_s$ (Ω*cm$^2$) | $R_{sh}$ (Ω*cm$^2$) |
|---|---|---|---|---|---|---|
| $Sb_2Se_3$+ $TiO_2$[18] | $9.065\times10^{-6}$ | 0.7675 | $1.496\times10^{-7}$ | $9.573\times10^{-12}$ | 311.6 | $1.876\times10^{13}$ |
| $Sb_2Se_3$+ $TiO_2$+Co&Cu[18] | $1.000\times10^{-4}$ | 0.9895 | $1.354\times10^{-3}$ | $1.124\times10^{-23}$ | 114.7 | $3.370\times10^{13}$ |

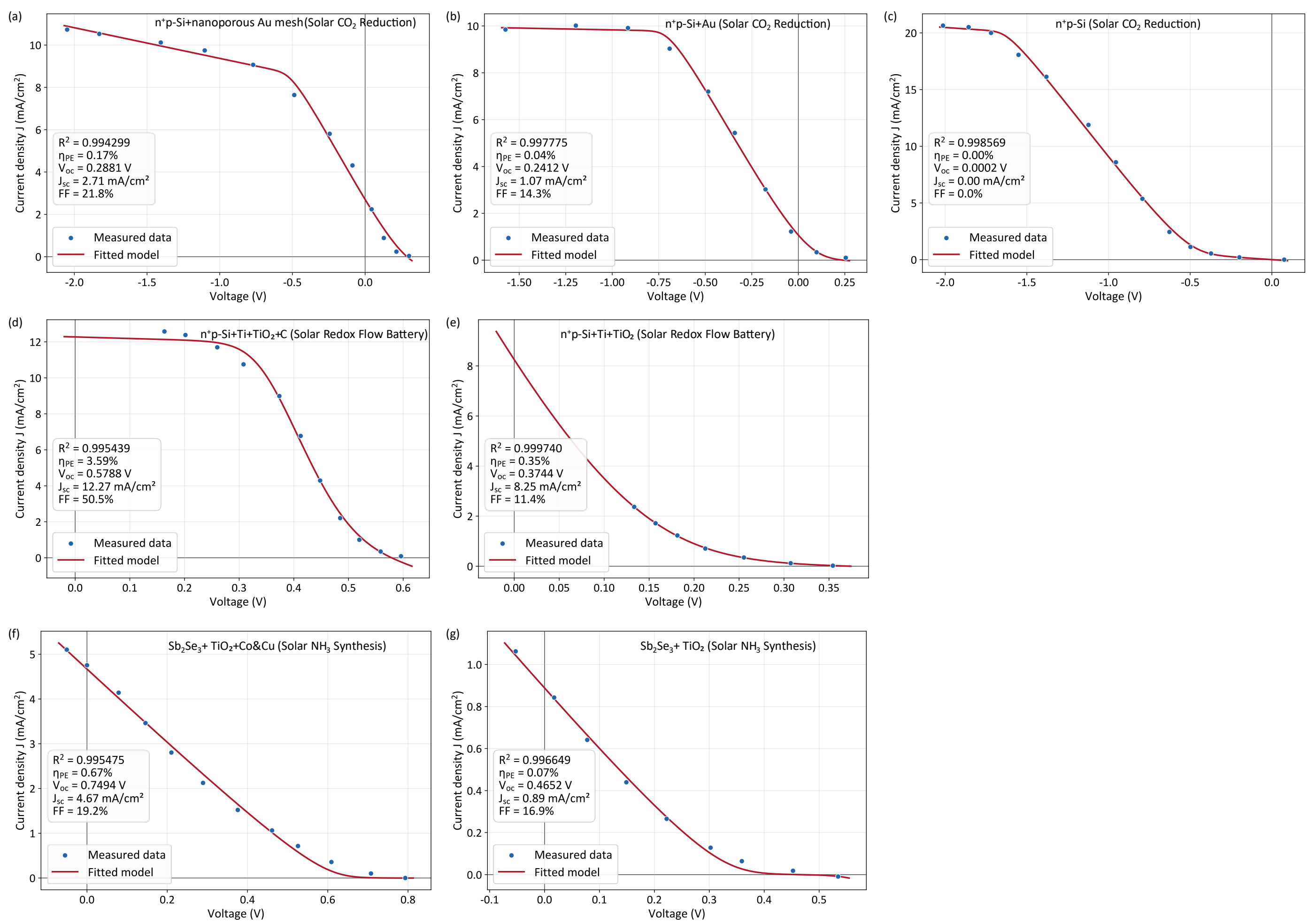

**Figure S6. J-V Curve Fitting of Photoelectrodes for Solar $CO_2$ Reduction, Solar Redox Flow Batteries, and Solar Ammonia Synthesis.**

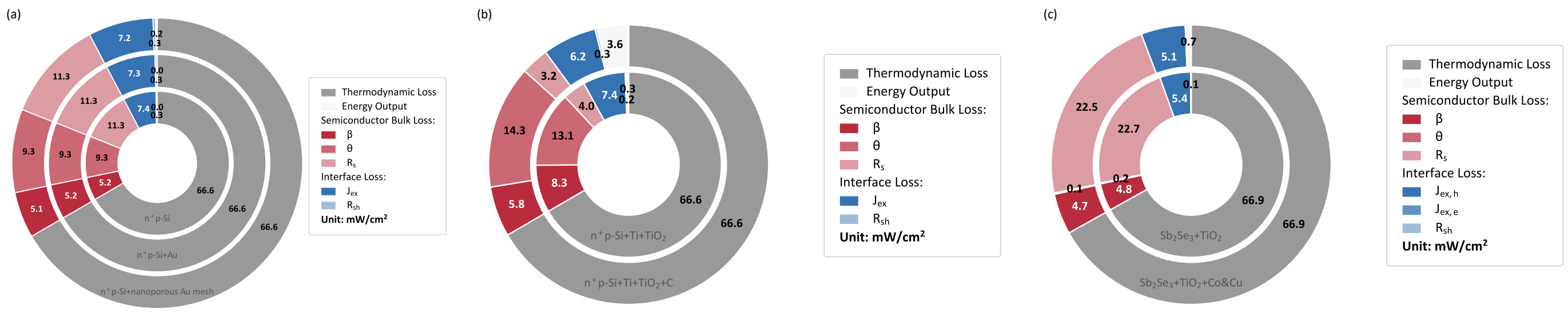

**Figure S7. Donut chart visualization of energy loss distribution in photoelectrodes for solar $CO_2$ reduction, solar redox flow batteries, and solar ammonia synthesis.**

# 8. Multidimensional Evaluation Framework for Photoelectrode Materials and Literature Selection

A multidimensional evaluation framework was constructed in this work for the comparison of photoelectrode materials, incorporating five key metrics: elemental abundance, operational stability, actual efficiency, theoretical efficiency, and toxicity.

Elemental abundance is defined by the crustal abundance of the rarest element in the material composition, as listed in Table S8. For improved visualization and comparability in the radar chart, the abundance values in mg per kg are transformed using a base 10 logarithm prior to plotting.

Specifically, for a given material, taking 70% of the highest photocurrent density of the previously selected work as the threshold, the one with the highest stability is selected from all works with photocurrent densities above this threshold to represent the stability of the material, as summarized in Table S8.

The efficiency metrics include the actual efficiency and the theoretical efficiency. The former refers to the efficiency corresponding to the photoelectrode using the nine representative materials with the highest reported photocurrent density up to the end of 2025, while the latter represents the theoretical limit. Their specific definitions have been detailed in the main text.

The toxicity metric is evaluated based on the highest acute toxicity category of the material and its precursors, with information obtained from chemical manufacturers' datasheets, as listed in Table S8. For example, when the Hazard Classification is labeled as "Acute Tox. 4 Oral", the corresponding toxicity level is 4, indicating that the oral median lethal dose (LD50) falls within the range of 300-2000 mg/kg. If no corresponding hazard classification is available, the material is considered to be essentially non-toxic, and the toxicity level is assigned a value of 5. For perovskite materials, the toxicity evaluation is based on tin lead halide perovskites, as these correspond to the compositions used in the highest photocurrent experimental studies considered in this work.

**Table S8. Elemental Abundance, Stability, and Toxicity of Photoelectrode Materials and Their Corresponding Data Sources**

| Photoelectrode material | Abundance(mg/kg)[19] | Stability (h/1% decay) | Toxicity |
|---|---|---|---|
| $Cu_2O$ | 60 (Cu) | 12.50[14] | 4[20] |
| α-$Fe_2O_3$ | 56300 (Fe) | 35.00[21] | 5 |
| $Ta_3N_5$ | 2 (Ta) | 12.00[13] | 4[22] |
| $BiVO_4$ | 0.0085 (Bi) | 46.67[23] | 3[24–26] |
| Si | 282000 (Si) | 86.96[27] | 5 |
| Tin-Lead Halide Perovskite | 0.45 (I) | 140.00[28] | 4[29,30] |
| CIGS | 0.05 (Se) | 0.67[9] | 4[31–33] |
| CZTS | 0.05 (Se) | 8.89[34] | 4[31,35–41] |
| $Sb_2Se_3$ | 0.05 (Se) | 3.33[42] | 3[43] |